\documentclass[12pt,preprint]{aastex}
\usepackage{amssymb,amsmath,color,hyperref}
\newcounter{address}
\renewcommand{\theaddress}{\alph{address}}
\newcounter{tableone}
\definecolor{linkcolor}{rgb}{0,0,0.25}
\hypersetup{
  colorlinks=true,        
  linkcolor=linkcolor,    
  citecolor=linkcolor,    
  filecolor=linkcolor,    
  urlcolor=linkcolor      
}
\newcommand{\etal}{et~al.}
\newcommand{\eg}{e.g.}
\newcommand{\ie}{i.e.}
\newcommand{\unit}[1]{\mathrm{#1}}
\newcommand{\AU}{\unit{AU}}

\newcommand{\yr}{\unit{yr}}
\newcommand{\satellite}[1]{\textsl{#1}}
\newcommand{\Gaia}{\satellite{Gaia}}
\newcommand{\KS}{K--S}
\newcommand{\figurenames}{\figurename s}
\newcommand{\sectionname}{\S}
\newcommand{\tvector}[1]{\boldsymbol{\vec{#1}}}
\newcommand{\vx}{\tvector{x}}
\newcommand{\vv}{\tvector{v}}
\newcommand{\va}{\tvector{a}}

\newcommand{\tuvector}[1]{\boldsymbol{\hat{#1}}}
\newcommand{\rhat}{\tuvector{r}}
\newcommand{\mvector}[1]{\boldsymbol{{#1}}}
\newcommand{\mvtheta}{\mvector{\theta}}
\newcommand{\mvthetae}{\mvector{\theta}_e}
\newcommand{\mvthetaepsilon}{\mvector{\theta}_{\epsilon}}
\newcommand{\mvomega}{\mvector{\omega}}

\newcommand{\dd}{\mathrm{d}}
\newcommand{\setofxv}{\{\vx_i,\vv_i\}}

\newcommand{\ueff}{u_{\mathrm{eff}}}
\newcommand{\rperi}{r_{\mathrm{peri}}}
\newcommand{\rap}{r_{\mathrm{ap}}}
\renewcommand{\angle}{\phi_r}
\newcommand{\anglei}{\phi_{r,i}}

\newcommand{\lnepsilon}{\ln{\epsilon}}
\newcommand{\vperp}{j^2} 
\newcommand{\vperpi}{j_i^2}

\newcommand{\eqnname}{equation}

\begin{document}

\title{
Dynamical inference from a kinematic snapshot:\\
  The force law in the Solar System}
\author{
  Jo~Bovy\altaffilmark{\ref{CCPP},\ref{email}},
  Iain~Murray\altaffilmark{\ref{Toronto}},
  David~W.~Hogg\altaffilmark{\ref{CCPP},\ref{MPIA}}
}
\altaffiltext{\theaddress}{\label{CCPP}\refstepcounter{address}
  Center for Cosmology and Particle Physics, Department of Physics,
  New York University, 4 Washington Place, New York, NY 10003, USA}
\altaffiltext{\theaddress}{\label{email}\refstepcounter{address}
  Correspondence should be addressed to jo.bovy@nyu.edu~.}
\altaffiltext{\theaddress}{\label{Toronto}\refstepcounter{address}
  Department of Computer Science,
  University of Toronto, Toronto, Ontario M5S 3G4, Canada}
\altaffiltext{\theaddress}{\label{MPIA}\refstepcounter{address}
  Max-Planck-Institut f\"ur Astronomie,
  K\"onigstuhl 17, D-69117 Heidelberg, Germany}

\begin{abstract}
If a dynamical system is long-lived and non-resonant (that is, if
there is a set of tracers that have evolved independently through many
orbital times), and if the system is observed at any non-special time,
it is possible to infer the dynamical properties of the system (such
as the gravitational force or acceleration law) from a snapshot of the
positions and velocities of the tracer population at a single moment
in time. In this paper we describe a general inference technique that
solves this problem while allowing (1)~the unknown distribution
function of the tracer population to be simultaneously inferred and
marginalized over, and (2)~prior information about the gravitational
field and distribution function to be taken into account. As an
example, we consider the simplest problem of this kind: We infer the
force law in the Solar System using only an instantaneous kinematic
snapshot (valid at 2009 April 1.0) for the eight major planets. We
consider purely radial acceleration laws of the form $a_r=
-A\,[r/r_0]^{-\alpha}$, where $r$ is the distance from the Sun.  Using
a probabilistic inference technique, we infer $1.989<\alpha<2.052$
(95~percent interval), largely independent of any assumptions about
the distribution of energies and eccentricities in the system beyond
the assumption that the system is phase-mixed. Generalizations of the
methods used here will permit, among other things, inference of
Milky Way dynamics from \Gaia-like observations.
\end{abstract}


\section{Introduction}

The \Gaia~Satellite \citep{Perryman01a} will measure positions and
velocities for millions to billions of stars at varying precision.
One of the principal goals of this mission is to provide the data
necessary to infer the dynamical state of the Milky Way.  However,
there are issues in principle with inference of dynamics from a
\emph{snapshot} or instantaneous set of configuration and velocity
measurements: The instantaneous positions and velocities have no
\emph{necessary} relationship with the gravitational potential or
accelerations.  Indeed, despite a considerable literature \citep[for
  example,][]{Oort32, Schwarzschild79, Little87a, Binney94,
  Johnston99, roulette} there is no methodology for performing this
inference that naturally handles all of the issues, including finite
and non-trivial observational uncertainties or noise, missing data,
non-steady aspects of the mass distribution, and the (incredibly
likely) possibility that the potential is not (simply) integrable.
Robust inference may not even be \emph{possible} if the Milky Way has
significant time-dependence or is strongly chaotic or is far from
showing any simple symmetries (such as axisymmetry).

Certainly there is no hope for dynamical inference on the massive
scale required for the \Gaia\ data set if we cannot perform it on much
simpler, much more symmetrical, much older (in a dynamical sense), and
much smaller (in a data sense) systems.  In what follows, we take one
of the simplest possible systems---the Solar System---and the smallest
possible data set---the positions and velocities of the major planets
at a moment in time---and perform a complete dynamical inference.  For
a test system we could also have chosen the black hole at the Galactic
Center, where similar considerations apply.  However, this system has
additional issues with ``missing data'' because not all six
phase-space coordinates are directly measurable for all stars.  The
Solar System truly is the simplest problem in this class.

Our inferential starting point is orbital roulette \citep{roulette}.
In this previous work, it was assumed that the orbital angles (in the
action--angle formalism) are uniformly distributed. Thus, dynamical
model parameters that correspond to orbital angles that suspiciously
lack diversity are rejected. More specifically, the method considers a
large range of possible dynamical model parameters, computes orbital
angles at every value of the parameters, and a distribution statistic
on those orbital angles. If the distribution statistic is designed to
monotonically increase with the diversity of the angles or the
flatness of the angle distribution, the ``best fit'' dynamical model
parameters are those that optimize the distribution statistic.  This
kind of approach is inherently frequentist: on many applications of
these procedures to independent datasets a confidence interval
captures the true dynamical parameters on a guaranteed fraction of
trials. However, for any given dataset the confidence interval
produced might not represent a credible set of parameters.

In what follows we cast the dynamical parameters estimation as a
probabilistic inference problem (a ``Bayesian'' approach). We adopt
the same core assumption as the roulette problem: we assume that the
orbital angles are uniformly distributed. However, in this framework,
we must also construct prior probability distribution functions for
the dynamical model parameters and conditional probabilities of the
data given the model to encode the assumptions of a long-lived and
bound Solar System.  These prior and conditional probability
distributions and the data create posterior probability distribution
functions for all the parameters. We use this new method to infer the
gravitational force law (radial dependence and amplitude).  What is
new here, in the context of Solar System dynamics, is that we perform
this inference with only a snapshot of the kinematic state, that is,
with only the positions and velocities of the planets at a
\emph{single instant of time.}

Of course the kinematic snapshot we employ is, in fact, a set of
initial conditions for a Solar System integration \citep{Giorgini96a}.
These initial conditions were determined not by a single measurement
at a single epoch, but are in fact the result of an optimization of a
Solar System integration to observed planetary positions over many
decades.  In the context of this paper---a demonstration of a
method---it is best to think of these ``data'' as ``simulated data''
useful for testing the method.  They just happen to be data that have
been simulated by the analog computer we know as the Solar System.

What is new here, in the context of dynamical inference, is that our
method is fully probabilistic or Bayesian.  This is important for
future problems, such as the \Gaia\ problem, or for inferring the mass
of the black hole at the center of the Milky Way, because in these
real data analysis problems, the data points come with non-trivial and
highly correlated observational uncertainties, and because entire
dimensions of phase space are missing or unobserved.  At the Galactic
Center, we do not know the radial distance to anything accurately, and
in the \Gaia\ data set many of the radial velocities will not be
measured.  The Bayesian framework handles these real-world data issues
very naturally, although in fact they are \emph{not} important in the
test problem we solve here.  Aside from these issues of principle, it
is also the case, as we will show, that the Bayesian method performs
extremely well.

Of course, a lot is known about the gravitational force law in the
Solar System, so we do not expect, at the outset, to be surprised by
our results.  The first force-law inference in the Solar System
\citep[][and also work by contemporaries, particularly Hooke, who may
  have priority]{Newton} made use of full orbit shape determination
\citep{Kepler}.  In this sense, Newton's problem---find the force law
(from among a small, discrete group of possible force laws) that leads
to elliptical orbits with the Sun at one focus---was much easier than
the problem we have set for ourselves.  Of course, along the way,
Newton had to develop for the first time the general principles of
kinematics and dynamics!

\section{Parameterized force law or dynamical model}\label{sec:model}

We are going to assume spherical symmetry of the Solar System's force
law and gravitational potential, although nothing in the general
inference formalism that follows will require this.  Consider a radial
force law (really acceleration law) of the form
\begin{equation}
\va = -A\,\left[\frac{r}{r_0}\right]^{-\alpha}\,\rhat \quad,
\end{equation}
where $A$ is an amplitude, $r$ is the distance from the Sun, $r_0$ is
a distance scale (in this case we will use $r_0=1\,\AU$ so that $A$
can be thought of as the acceleration at Earth's orbit), $\alpha$
parameterizes the radial dependence, and $\rhat$ is the radial
direction.  In this model, the list of free parameters is
\begin{equation}
\mvomega \equiv \{\ln A,\alpha\}\quad,
\end{equation}
where we have taken the logarithm of $A$ because in inference
problems, dimensioned parameters are usually best handled in the log
\citep{Jeffreys39a,Sivia06a}.

The potential $u$ (potential energy per unit planet mass), radial
effective potential $\ueff$, and binding
energy per unit mass $\epsilon\equiv -E/m$ are
\begin{equation}
u(r) = \frac{A\,r_0}{1-\alpha}\,\left[\frac{r}{r_0}\right]^{1-\alpha} \quad,
\end{equation}
\begin{equation}
\ueff(r) = u(r) + \frac{j^2}{2\,r^2} \quad,
\end{equation}
\begin{equation}
\epsilon = - \ueff - \frac{1}{2}\,v_r^2 \quad,
\end{equation}
where $j^2$ is the square of the magnitude of the planet's angular
momentum per unit mass (or $L^2/m^2$), and $v_r$ is the radial
component of the velocity (the component of $\vv$ parallel to $\vx$).
The perihelion and aphelion distances
$\rperi$ and $\rap$ are both found by setting $\epsilon=-\ueff$.  With
these, we can define a radial asymmetry $e$ as
\begin{equation}
e \equiv \frac{\rap - \rperi}{\rap + \rperi} \quad,
\end{equation}
where we have called this ``$e$'' because in the Kepler--Newton
$\alpha=2$ case it is the orbital eccentricity.  One way of thinking of
this radial asymmetry is that at any point in the space made up of the
dynamical parameters $\mvomega$ and the binding energy $\epsilon$,
the radial asymmetry $e$ is a dimensionless description of the
angular momentum magnitude.

Importantly for what follows, we can define a ``radial angle''
$\angle$ that increases linearly with time from perihelion passage
through next perihelion passage.  Any planet at radius $r$ on an orbit
with perihelion distance $\rperi$ and aphelion distance $\rap$ can be
assigned this angle $\angle$ by
\begin{equation}
\angle \equiv \left\{\begin{array}{cl}\displaystyle
  \pi\,\frac{t(r)-t(\rperi)}{t(\rap)-t(\rperi)} & \mbox{for $v_r > 0$}
  \\[2.5ex] \displaystyle
  \pi+\pi\,\frac{t(r)-t(\rap)}{t(\rperi)-t(\rap)} & \mbox{for $v_r < 0$}
\end{array}\right. \quad,
\end{equation}
where the first numerator is the time it takes to go from $\rperi$ to
$r$ outbound, the first denominator is the time it takes to go from
$\rperi$ to $\rap$ outbound, the second numerator and denominator are
the times inbound, and all time differences between two radii can be
computed numerically for general values of $\alpha$ by integrating the
inverse of the radial velocity between these radii. The first-order
form of this integral has an integrable singularity at the perihelion
and aphelion, which can be handled by an appropriate change of
variables \citep[\eg, ][]{Press07a}. A planet observed at a set of
random times spanning many orbits will be observed to have radial
angles $\angle$ drawn from a flat distribution in the range
$0<\angle<2\,\pi$.  This radial angle is one of the angles in the
action--angle formulation of the system, which is integrable for the
simple reason that it is spherically symmetric.

\section{Kinematic data}

In what follows, we are going to use and compare several methods for
inferring the force-law parameters $\mvomega$ (the amplitude $\ln A$
and radial exponent $\alpha$ of the spherical force law) from an
instantaneous snapshot of the positions and velocities of the eight
major planets.  This snapshot was taken from JPL's HORIZONS System
which provides highly accurate ephemerides for Solar System
objects\footnote{Available at \url{http://ssd.jpl.nasa.gov/?horizons}}
\citep{Giorgini96a}. It is an extrapolation (at the time of writing)
to 2009 April 1.0, approximately 400 years after the important
publication of Kepler (1609). This kinematic snapshot is given in
Table~\ref{table:eph}.

Since this snapshot is obtained by integrating the positions and
velocities of Solar System bodies, the accuracy is limited by (i) the
correctness of the dynamical model used, (ii) the numerical
integration of the equations of motion, and (iii) the accuracy to
which the initial conditions are known. It is generally believed that
the dynamical model used is correct and complete, and that the
numerical integration is sufficiently accurate. The main uncertainty
in the ephemerides is then that due to the uncertainty in the initial
conditions. The current set of initial conditions
\citep[DE405;][]{Standish98a} is a fit to a set of optical, radar, and
VLBI observations as well as to a set of spacecraft range and Doppler
points from various space missions. The uncertainties are the largest
for the outer planets, since the data for these are almost entirely
from optical observations (with the exception of Jupiter), and because
Neptune has not been observed over a full orbit since the start of
precise measurements. A comparison between the DE405 ephemerides and
more recent observations shows that the positions of the inner planets
are known to a fractional accuracy of approximately $10^{-8}$, while
those of the outer planets are known to a fractional accuracy of
$10^{-6}$ to $10^{-7}$ \citep{Standish04a}. Uncertainties in the
velocities are at the same fractional magnitude.

This kinematic snapshot is not, of course, a fair data set with which
to perform the inference below, for the main reason that the
``measured'' kinematic state of the Solar System is in fact the output
of fitting observations with a dynamical model that \emph{assumes}
$\alpha=2$.  For this reason, the data should be thought of as
``idealized'' or ``simulated data'' and the work must be considered a
test of the method rather than a definitive inference.

\section{Bound, virialized, and long-lived}

\begin{figure}
\includegraphics[width=.75\textwidth]{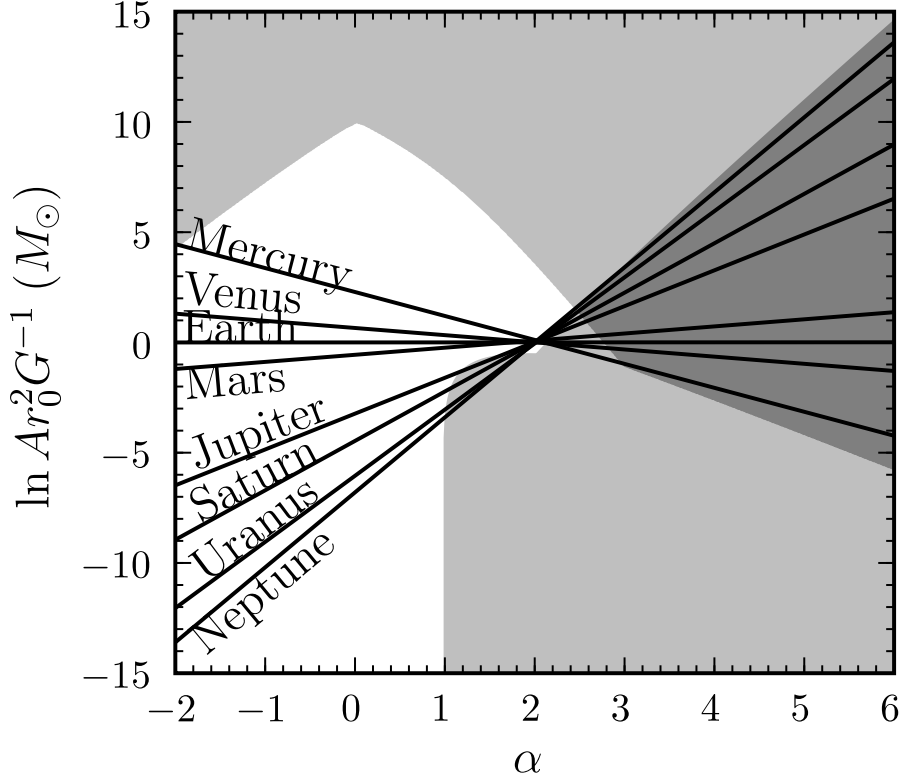}
\caption{The Virial relation between the kinetic energy and the
  potential energy (equation~[\ref{eq:virialtheorem}]) for each of the
  eight planets in the Solar System. For the combinations of dynamical
  parameters in the light gray region in the lower right at least one
  planet becomes unbound. When the dynamical parameters are in the
  light gray region in the upper left at least one planet has
  $\rperi<R_\odot$. The light gray regions overlap in the dark gray
  region. In the units used in this \figurename, the ``true'' value of
  $A$ lies at $\ln A r_0^2 G^{-1} = 0$.}\label{fig:virial_main}
\end{figure}

\begin{figure}
\includegraphics[width=.75\textwidth]{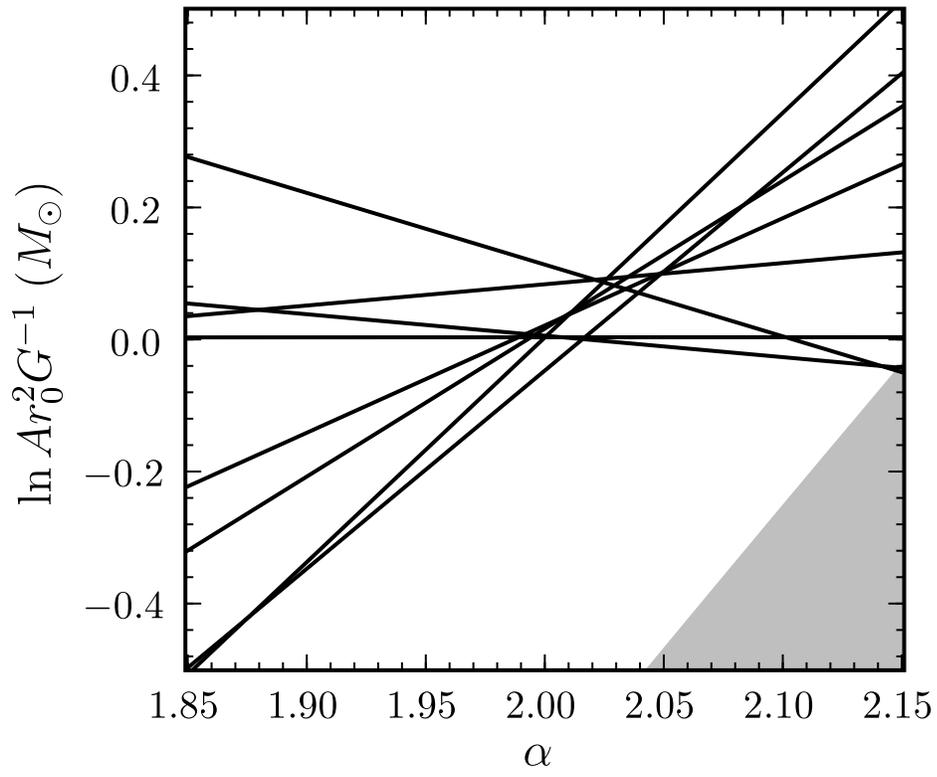}
\caption{Zoomed in version of \figurename~\ref{fig:virial_main}.}\label{fig:virial_zoom}
\end{figure}

The virial theorem relates the time averages $\left<T\right>$ and
$\left<U\right>$ of a test particle's kinetic and potential energies
through
\begin{equation}\label{eq:virialtheorem}
\left<T\right> = \frac{1-\alpha}{2}\,\left<U\right> \quad,
\end{equation}
where $\alpha$ is the exponent in the radial force law.  Given that a
planet's potential energy is a function of the dynamical parameters
$\mvomega$, while its kinetic energy is not, the virial relation for
each planet becomes a one-dimensional locus in the $\mvomega$ space.
Using kinetic and potential energies computed from the observations as
a proxy for the time-averaged energies, these loci are shown in
\figurenames~\ref{fig:virial_main} and \ref{fig:virial_zoom}.  The
fact that all eight lines cross near a single point in the space is
encouraging that the system is virialized (as we expect) and that the
inference will work. That the eight lines so nearly intersect in a
single point is because, in fact, many of the planets are on circular
orbits; the probabilistic method developed in this paper does not
assume this, but it does retain the precision that is possible because
of this situation.

Also shown in \figurenames~\ref{fig:virial_main} and
\ref{fig:virial_zoom} is the region of parameter space in which one or
more of the planets is unbound because $T>U$ or, equivalently,
$\epsilon < 0$.  In what follows, we will assign vanishing probability
to regions of parameter space in which one or more planets is
unbound. However, as we will see, this does not affect any of our
conclusions. Also shown in \figurename~\ref{fig:virial_main} is the
region of parameter space in which one or more of the planets has
$\rperi<R_\odot$, where $R_\odot$ is the radius of the Sun.  This part
of parameter space ought also to be excluded, although in practice
this is not necessary for any of what follows.

\begin{figure}
\includegraphics[height=.2\textheight]{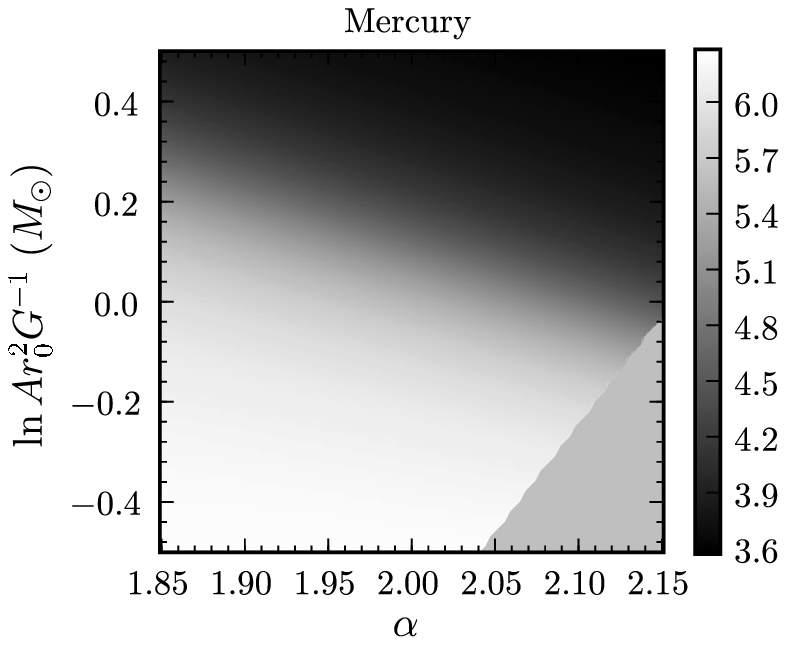}
\includegraphics[height=.2\textheight]{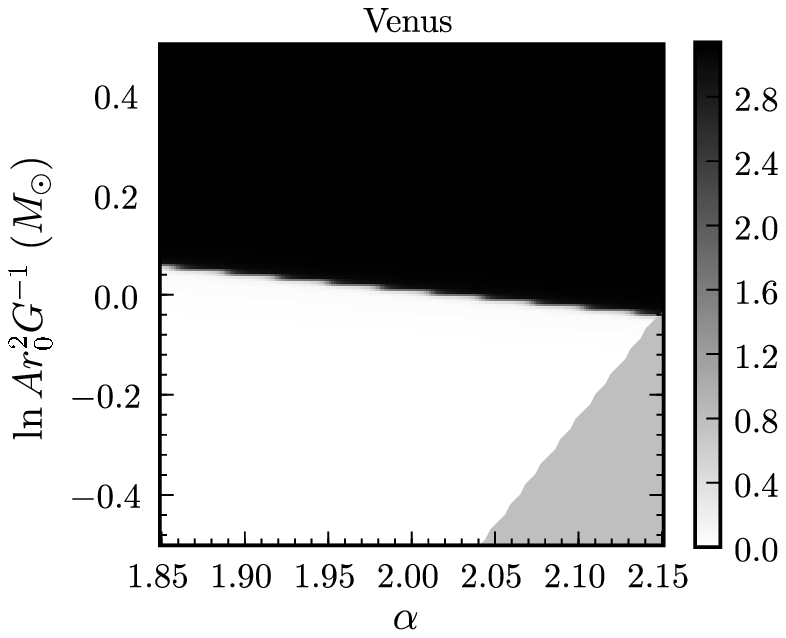}\\
\includegraphics[height=.2\textheight]{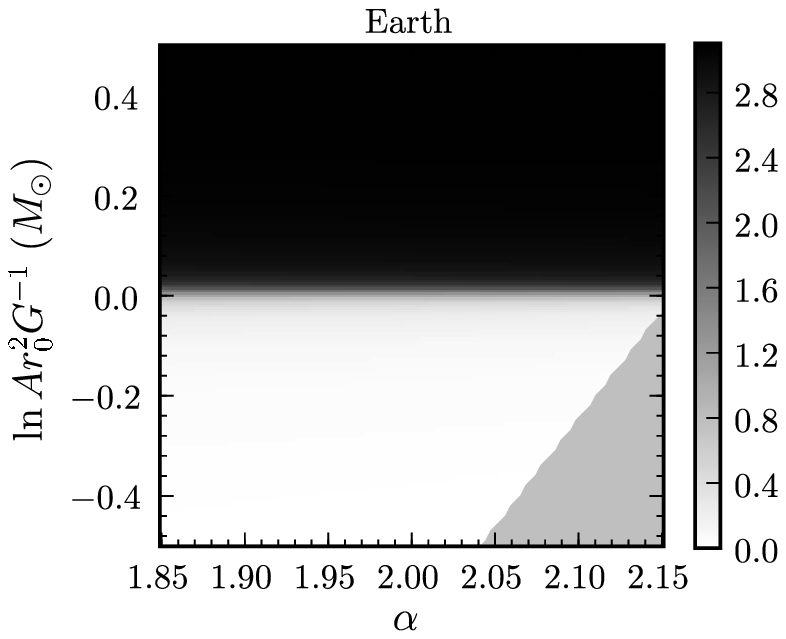}
\includegraphics[height=.2\textheight]{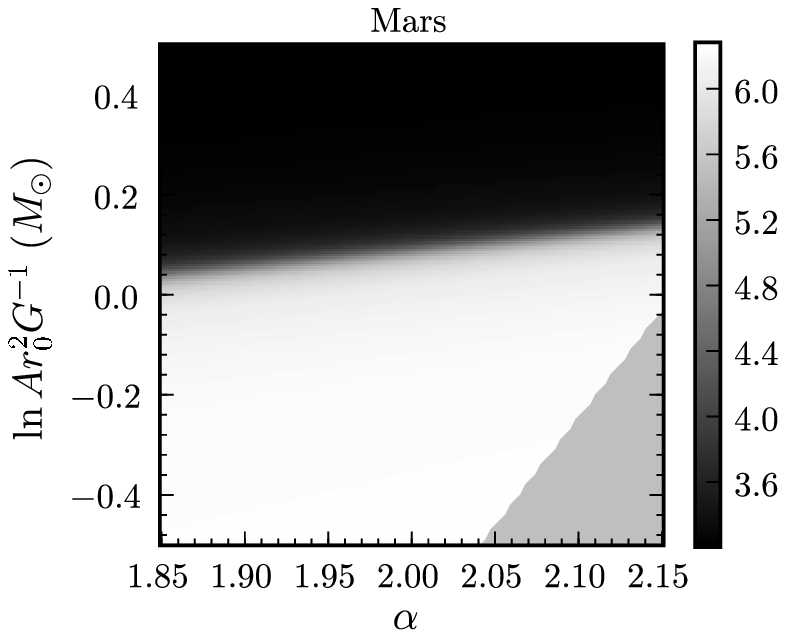}\\
\includegraphics[height=.2\textheight]{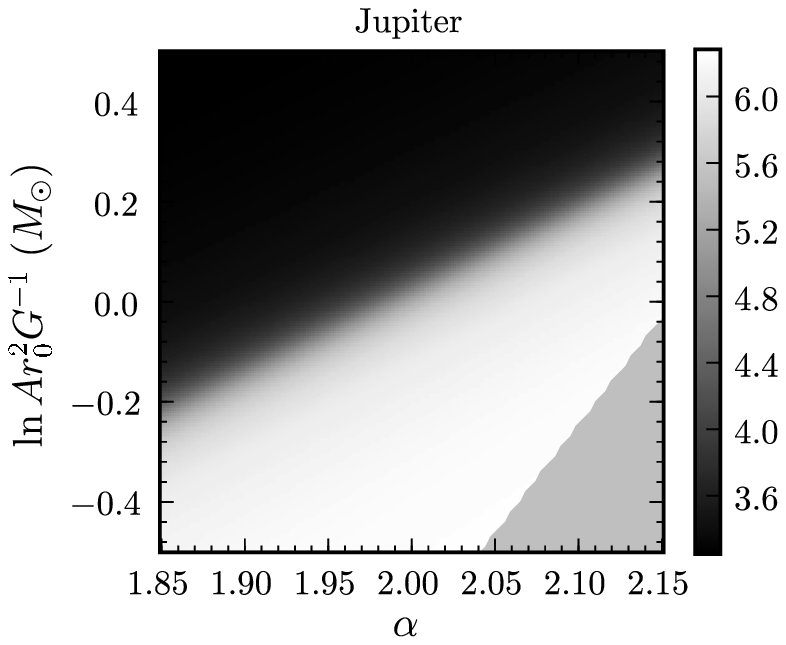}
\includegraphics[height=.2\textheight]{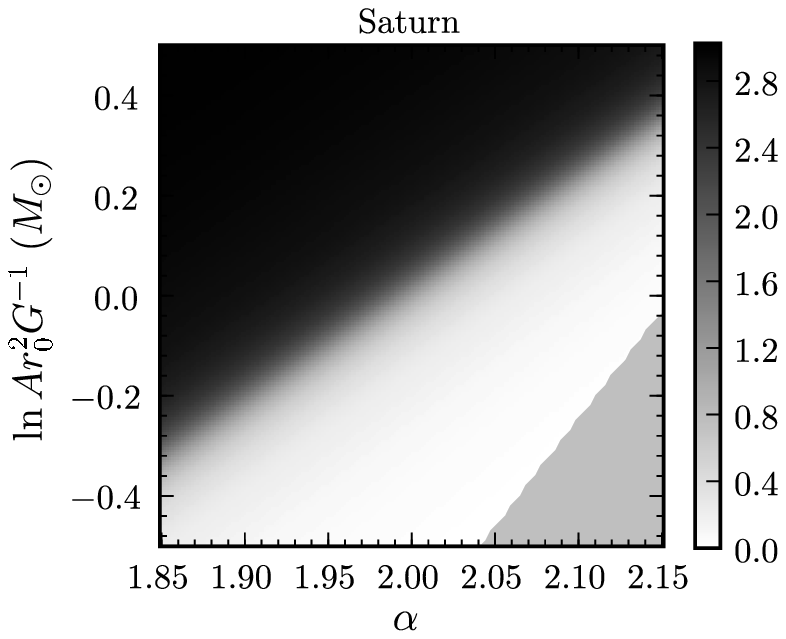}\\
\includegraphics[height=.2\textheight]{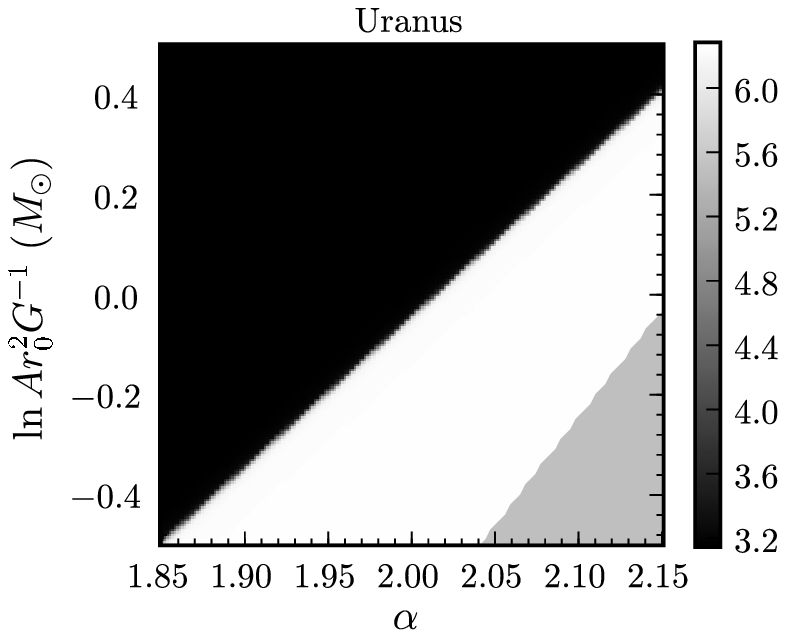}
\includegraphics[height=.2\textheight]{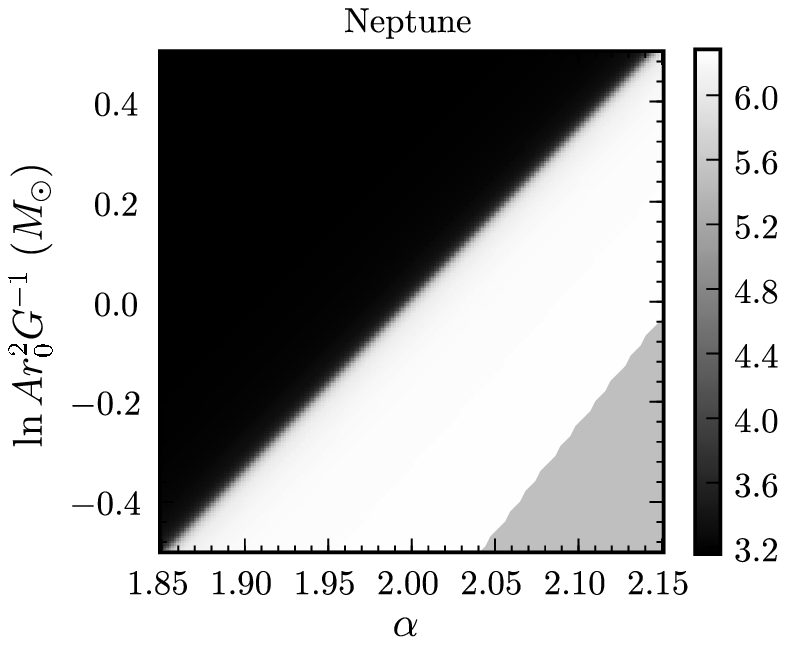}
\caption{Computed radial angles $\angle$ for each of the eight planets
  as a function of the dynamical parameters. The gray triangular
  region in the bottom-right corner is the region excluded by the
  condition that all the planets are bound.  Each planet has an angle
  range of $0<\angle<\pi$ if it has radial velocity $v_r>0$ (outgoing
  from perihelion) or $\pi<\angle<2\,\pi$ if it has $v_r<0$ (incoming
  from aphelion).}\label{fig:anglesPlanets}
\end{figure}

In preparation for what follows, we pre-compute all of the planet
radial angles $\anglei$---given their positions $\vx_i$ and velocities
$\vv_i$---as a function of the dynamical parameters.  These angles are
shown in \figurename~\ref{fig:anglesPlanets}.

\section{Frequentist orbital roulette}\label{sec:freq}

In orbital roulette \citep{roulette}, the idea is to compute, for each
point $\mvomega$ in parameter space, the $N$ radial angles $\anglei$
and analyze them statistically for being well mixed.  In practice,
this means applying a distribution test or multiple distribution tests
to the angles and preferring parameters for which these tests are more
consistent with a random or flat distribution of angles in the range
$0<\angle<2\,\pi$.  Because each test provides one constraint, and in
the case described here the parameter space is two-dimensional, at
least two qualitatively different tests are required to make localized
constraints in $\mvomega$.  In addition to a test for a flat angle
distribution we also apply a test for an angle--energy correlation.

About the simplest consistency test for the calculated angles is a
test of the mean of the angles: Is the mean consistent with the
expected mean for a uniformly distributed set of $N$ planets? For this
to perform well one must fold the angles of the inbound planets onto
the interval $[0,\pi]$, that is, disregard the information in the
\emph{sign} of the radial velocity; then the expected mean of the
angles is equal to $\pi/2$ (for details on how to test this
assumption, see \citealt{roulette}). The fact that we have to perform
this mapping indicates that the procedure is \textit{ad
  hoc}. Indeed, for a uniform distribution on the circle there is no
specific meaning to the perihelion and the aphelion, or to any two
points, such that no real meaning can be attached to the mean of the
angles between two arbitrary points on the circle.

Better, one can test the consistency of the full distribution function
of the angles with a uniform distribution. This could again be done
with the full $[0,2\pi]$ distributed angles or with the folded angles;
the results will depend on this choice. Testing an observed
distribution for consistency with an expected distribution often
involves comparing cumulative distribution functions
\citep{Kolmogorov41a}.  The Kolmogorov--Smirnov (\KS) test is the
simplest in practice, since the distribution of the test
statistic---the maximum difference between the cumulative
distributions---can be approximated by an analytic function
\citep{Stephens70a}. The \KS\ test is by construction most sensitive
to deviations near the median value; this rules out dynamical
parameters at which the planets bunch up at perihelion or at aphelion,
the situation in which about half are at perihelion and half at
aphelion can easily dupe the test.

A statistic can be chosen that is
sensitive to deviations at all values, such as the Anderson--Darling
statistic \citep{AndersonDarling}. However, no approximate analytic
description of the distribution of this statistic exists and in
practice this distribution has to be obtained by Monte Carlo sampling
\citep[\eg, ][]{roulette}. A statistic
more appropriate to the problem at hand (although we
are not primarily interested in a careful examination of the differences
between different frequentist procedures) is Kuiper's
statistic \citep{Kuiper62a}. This statistic---the sum of
the maximum distance of the observed cumulative distribution above and
below the expected cumulative distribution---is invariant under
periodic shifts of the angle and was specifically designed to test
uniform distributions on the circle. The advantage over
Anderson--Darling is that the asymptotic distribution of the
Kuiper statistic is known \citep[\eg, ][]{Press07a}.

\begin{figure}
\includegraphics[height=.2\textheight]{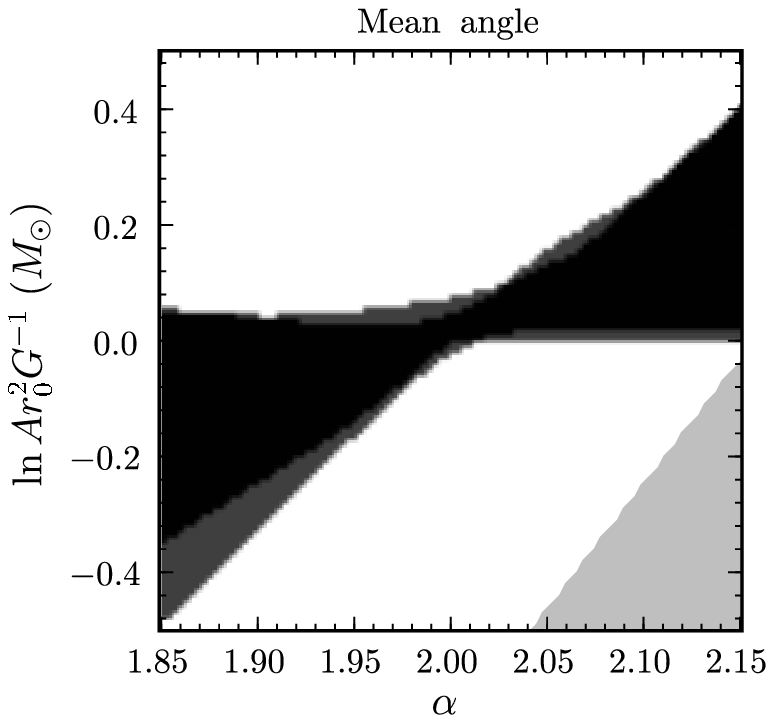}
\includegraphics[height=.2\textheight]{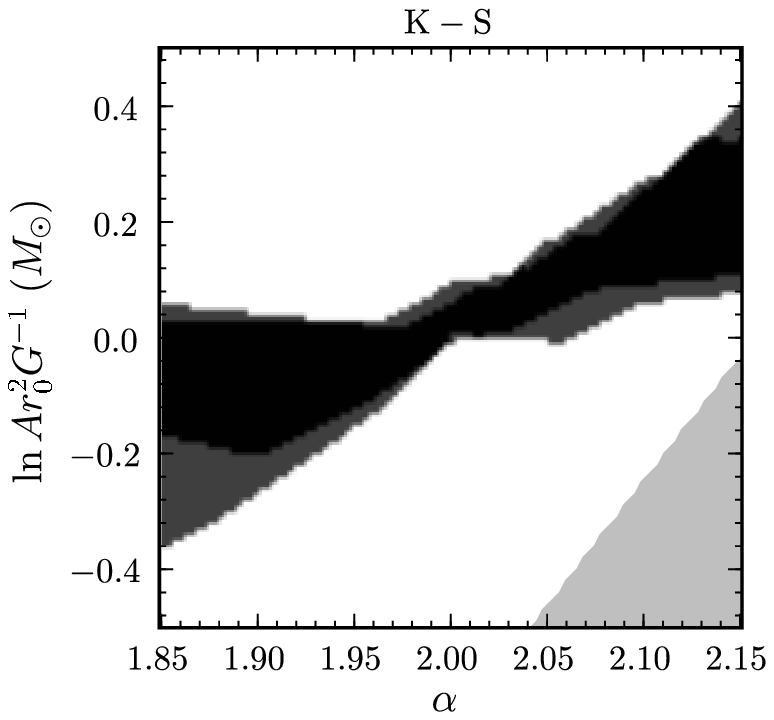}\\[5pt]
\includegraphics[height=.2\textheight]{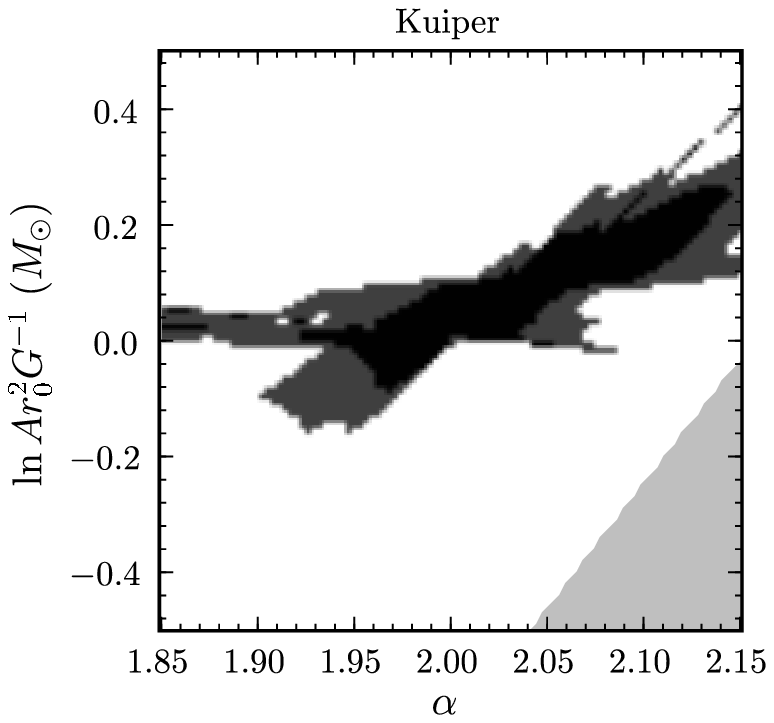}
\includegraphics[height=.2\textheight]{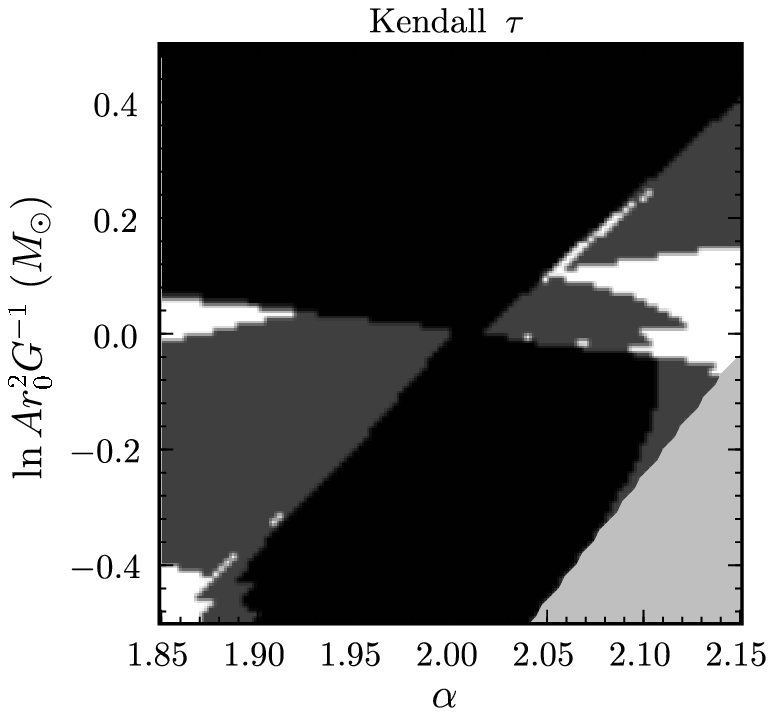}\\[5pt]
\includegraphics[height=.2\textheight]{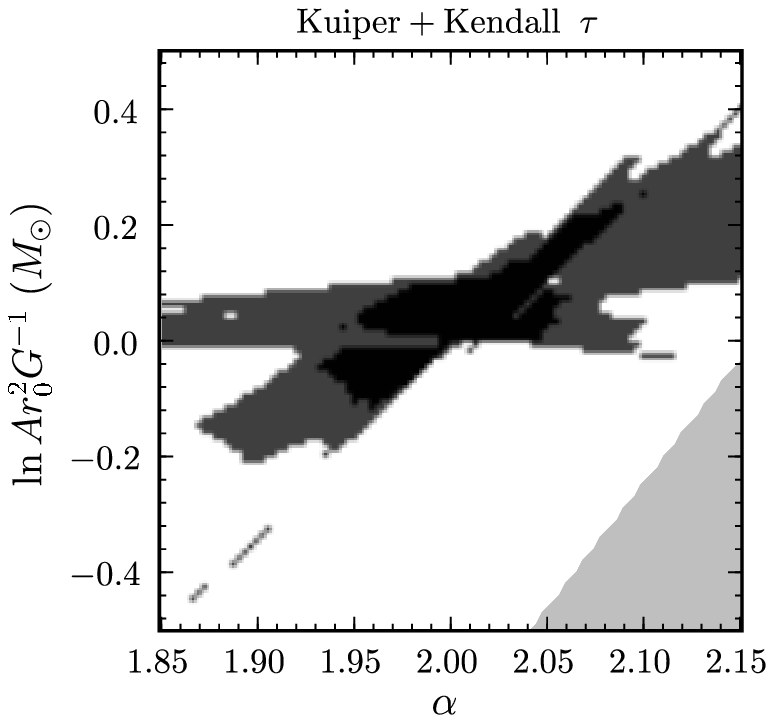}
\includegraphics[height=.2\textheight]{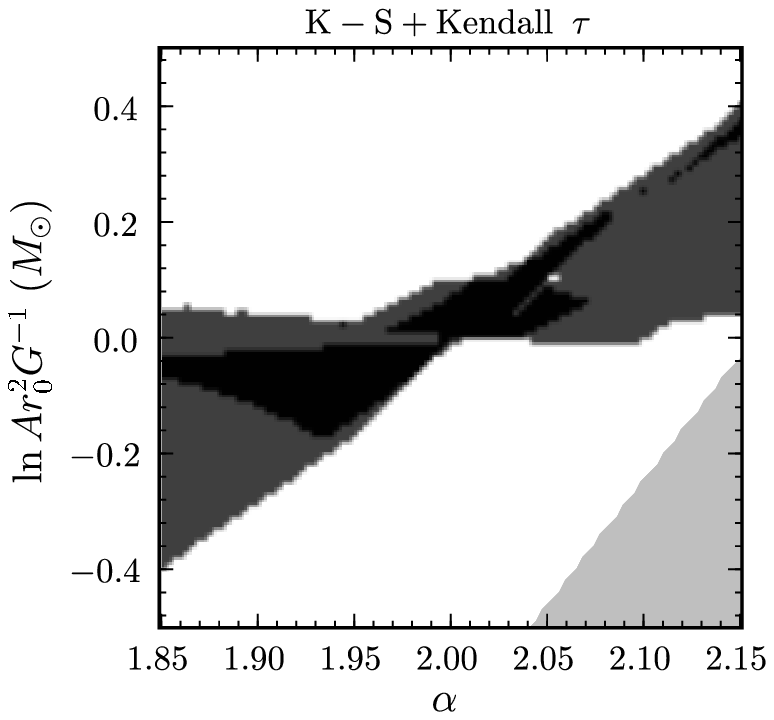}
\caption{Various frequentist tests applied to test the uniformity of
  the angle distribution and the absence of angle--energy
  correlations. From top to bottom, left to right: test of the mean of
  the angles; \KS\ test for the uniformity of the angle distribution;
  Kuiper test for the uniformity of the angles; Kendall $\tau$ test
  for the absence of angle--energy correlations; combined confidence
  intervals from the Kuiper test and the Kendall test; combined
  confidence intervals from the \KS\ test and the Kendall test. In the
  plots with a single statistic the darkest region corresponds to the
  95~percent confidence region, the lighter region to the 99~percent
  confidence region. The same is true for the plots with combinations
  of statistics, except that a Bonferroni correction has been applied
  to the significances of the individual statistics that make up the
  combination. In each plot the lightest region is excluded because at
  least one planet becomes unbound for those parameter
  values.}\label{fig:freq}
\end{figure}

All of these tests for the uniformity of the distribution of the angles
are shown in \figurename~\ref{fig:freq}. These tests can fail when for
a certain combination of dynamical parameters some planets are near
aphelion while other planets are near perihelion. This situation
appears here in \figurename~\ref{fig:anglesPlanets},
at $\alpha$ far from $2$, where there are
large regions in which the inner planets, especially Mercury and
Venus, are near perihelion, while the outer planets, especially Uranus
and Neptune, are near aphelion and \textit{vice versa}.  This prevents
the mean angle and \KS\ tests from excluding those regions. The Kuiper
test performs better.

A second constraint (for the two-dimensional parameter space) comes
from a second test. In regions of parameter space in which the inner
planets are all near perihelion and the outer planets are all near
aphelion, a significant correlation between the angles and the
energies exists. This correlation is unphysical if the system is not
being observed at any special time.  A non-parametric test for the
correlation is preferred here as the angle--energy correlation will
not in general be linear. We perform a test of the angle--energy
correlation using Kendall's $\tau$ \citep{Kendall38a}.  This is a rank
test; it only considers the relative ordering of the angles and
energies of different planets (for details on this test see
\citealt{Press07a}). That this test is in a sense orthogonal to the
tests of the uniformity of the angle distribution can be seen in
\figurename~\ref{fig:freq}.

All of the frequentist tests permit acceptance or rejection of a
dynamical model at a certain confidence level. 95 and 99~percent
confidence intervals for all of the frequentist tests are shown in
\figurename~\ref{fig:freq}. Also shown is the combination of the tests
of the uniformity of the angle distribution with the test of the
angle--energy correlation.

\section{Probabilistic dynamical inference}

The frequentist procedures perform reasonably well in this simple
problem because the number of dynamical parameters is small and the
data have vanishing errors and no missing components. As the number of
dynamical parameters increases the frequentist must find larger
numbers of tests in order to break degeneracies.  While data
uncertainties can be included by sampling the error distribution of
the data and combining the results \citep[\eg, ][]{roulette}, this
will perform badly in the limit of low signal-to-noise or missing
data. These difficulties are related to the fact that these procedures
use only a very crude model of the data, that is, that the angles are
distributed uniformly and that angle--energy correlations should be
absent, which allows no room for discovery of structure in phase
space. A fully Bayesian treatment of this problem can treat the
phase-space distribution function as an unknown function to be
inferred from the data. Modeling the full phase-space distribution
function permits simultaneous inference of missing data and properly
marginalized probability distributions for dynamical parameters.

Imagine that we have the three-vector positions $\vx_i$ and
three-vector velocities $\vv_i$ at some time $t$ for $N$ planets $i$
and a parameterized model for the gravitational acceleration law
(force law per unit mass) $\va_{\mvomega}(\vx)$, a function of
position $\vx$ and a list of parameters $\mvomega$.  We wish to obtain
an estimate of the posterior probability distribution
$p(\mvomega|\setofxv)$ for the parameters, where $\setofxv$ is the set
of all planet positions and velocities.  We employ Bayes's theorem as
follows:
\begin{equation}\label{eq:bayes}
p(\mvomega|\setofxv) = \frac{p(\setofxv|\mvomega)\,p(\mvomega)}{p(\setofxv)} \quad,
\end{equation}
where, as usual, $p(\setofxv|\mvomega)$ is the likelihood or the
probability distribution function for the data given the model
parameters, evaluated at the observed values of the data,
$p(\mvomega)$ is the prior probability distribution function for the
parameters, and the denominator is (for our purposes here) a
normalization constant.

For our chosen parameterization of the dynamical parameters,
$\mvomega$, a broad flat or uniform prior in the space represents a
reasonable description of our (assumed) prior knowledge.  The much
more challenging problem is to specify the likelihood of the dynamical
parameters, the conditional probability distribution function
$p(\setofxv|\mvomega)$. Without detailed knowledge of how the Solar
System formed, this probability distribution is also a representation
of our prior beliefs. Ideally our ability to learn from the observed
data will not be too sensitive to these beliefs.

It is easier to express beliefs about the angle~$\phi_r$, radial
asymmetry~$e$, and binding energy~$\epsilon$ of each planet than its
position and velocity. This is because, given our assumption that the
planets constitute an angle-mixed population, Jeans's theorem
\citep{Jeans15a,binneytremaine} tells us that the distribution
function, which is proportional to the probability of observing a
planet at a certain locus in phase space, is only a function of the
integrals of the motion. We wish to assign zero probability to
dynamical parameters that lead to any of the computed binding energies
$\epsilon(\vx_i,\vv_i,\mvomega)$ being negative (because we defined
$\epsilon > 0$ to be bound). We would also like to express our prior
information or assumption that the system is long-lived (in units of
the dynamical time), that it is non-resonant, and that we are not
seeing the system at any special time, or that the radial angles
$\angle$ will be randomly distributed between $0$ and $2\,\pi$. In the
absence of any better information, we will try to be as agnostic as
possible about the \emph{actions} (conserved quantities) of the
planets but extremely confident that all radial angles
$0<\angle<2\,\pi$ are equally likely.

In the simple spherical or radial situation under consideration here,
in which there are no missing data, we can rewrite the likelihood as a
function of the planets' radial coordinates, as the orientation of the
orbit does not depend on the dynamical parameters:
\begin{equation}
p(\vx,\vv|\mvomega) \propto
p(r,v_r,\vperp|\mvomega) =
|J(\lnepsilon,e,\angle;r,v_r,\vperp)| \;p(\lnepsilon, e, \angle|\mvomega) \quad .
\end{equation}
where, again, we have gone to $\lnepsilon$ because dimensioned
parameters are usually best handled in the log, and
$J(\lnepsilon,e,\angle;r,v_r,\vperp)$ is the Jacobian matrix of all
the partial derivatives of $(\lnepsilon,e,\angle)$ with respect to
$(r,v_r,\vperp)$. For spherical potentials, this Jacobian is given by
\begin{equation}\label{eq:jacgeneral}
|J(\lnepsilon,e,\angle;r,v_r,\vperp)| = \frac{2\,\pi}{T_r}\,\frac{1}{\epsilon}\,\left|\frac{\partial e}{\partial \vperp}\right|,
\end{equation}
where the derivative is evaluated at the current location in phase
space and $T_r$ is the radial period, which depends on the dynamical
parameters and the integrals of the motion. For general $\alpha$ this
radial period can be computed numerically (see
\sectionname~\ref{sec:model}). The derivative of the radial asymmetry
with respect to the specific angular momentum squared can be written
in terms of the perihelion and aphelion distance as
\begin{equation}\label{eq:dedj2}
\frac{\partial e}{\partial \vperp} = A r_0^{\alpha}\,\rap\, \rperi
\frac{\rap^{3-\alpha} - \rperi^{3-\alpha}}{(\rap+\rperi)^2}\,
\left[\vperp-A r_0^3\left(\frac{\rap}{r_0}\right)^{3-\alpha}\right]^{-1}
\left[\vperp-A r_0^3\left(\frac{\rperi}{r_0}\right)^{3-\alpha}\right]^{-1}\,.
\end{equation}
In the special case $\alpha = 2$ this Jacobian is
\begin{equation}\label{eq:jacobian}
|J(\lnepsilon,e,\angle;r,v_r,\vperp)| \propto \epsilon^{3/2}\,e^{-1}\,M_{\odot}^{-3}\,
\end{equation}
where we have dropped any terms that do not depend on the mass of the
Sun.

As well as assuming that the angles are independently and uniformly
distributed, we might further assume that the energy and radial
asymmetry of each planet were drawn independently:
\begin{equation}
    p(\{\lnepsilon_i, e_i, \anglei\}|\mvomega) =
    \frac{1}{(2\pi)^N}
    \prod_{i=1}^N p(e_i,\lnepsilon_i|\mvomega)\quad.
\end{equation}
However, a priori we do \emph{not} know the distribution function from
which radial asymmetries and binding energies were drawn. Fixing
$p(e_i,\lnepsilon_i|\mvomega)$ to a broad distribution would be making
a strong assumption: regardless of what the data say we would continue
to dogmatically believe that the distribution function was broad,
which could result in poor inferences about the model as a whole.

We can assume that the planets' properties were drawn independently
and identically distributed, without making strong assumptions about
what that distribution was. This is achieved by introducing auxiliary
parameters $\mvtheta = \{\mvthetae, \mvthetaepsilon\}$ that if known
would specify the distribution function. Since we do not know these
nuisance parameters, introducing a prior $p(\mvtheta|\mvomega)$ and
marginalizing over them,
\begin{equation}
    p(\{\lnepsilon_i, e_i, \anglei\}|\mvomega) = \frac{1}{(2\pi)^N}
    \int \dd\mvtheta \;\, p(\mvtheta|\mvomega) \prod_{i=1}^N p(e_i,\lnepsilon_i|\mvtheta,\mvomega)\quad,
\end{equation}
is then part of the inference task. With all of this in place, we
apply \eqnname~(\ref{eq:bayes}) to get the posterior distribution
over the dynamical parameters
\begin{equation}
p(\mvomega | \setofxv) \propto 
  \left[\prod_{i=1}^N |J(\lnepsilon_i,e_i,\anglei;r_i,v_{r,i},\vperpi)|\right] \,
  \int\dd\mvtheta\;p(\mvtheta,\mvomega)
  \,\prod_{i=1}^N
  \,p(e_i,\lnepsilon_i|\mvtheta,\mvomega)\quad ,
  \label{eqn:posterior}
\end{equation}
where each planet's value of $(\lnepsilon_i,e_i,\anglei)$ is a
function of phase-space position $(\vx_i,\vv_i)$ and dynamical
parameters $\mvomega$, and the Jacobian is evaluated at each planet's
value of $(\lnepsilon_i,e_i,\anglei)$.

For situations in which we have missing data or noisy observations,
further unknown quantities will be added to the model and marginalized
over. As the model becomes more complicated it is more likely that
Markov chain Monte Carlo \citep[\eg,][]{neal1993} or other approximate
computational methods will be required.

\subsection{Basic method}

To keep things as simple as possible, at first we model the
distribution function $p(\lnepsilon, e|\mvomega,\mvtheta)$ as a
product of a top-hat function in $\lnepsilon$ from $\lnepsilon_a$ to
$\lnepsilon_b$ with a top-hat function in $e$ from $e_a$ to $e_b$. In
this context, the phase-space parameter list is
\begin{equation}
\mvtheta = \{ \lnepsilon_a, \lnepsilon_b, e_a, e_b \} \quad.
\end{equation}
In what follows, we only consider values $A <\ln \epsilon_a < \ln
\epsilon_b < B$, where $A$ and $B$ provide very distant
(uninformative) limits (below, we will take the limit), and $0 \leq
e_a \leq e_b \leq 1$. These enforce our assumption or prior
information that the system is bound.

For our initial model we also set the prior $p(\mvtheta,\mvomega)$
flat or uniform in all the parameters. Now the marginalization
required to compute the posterior (equation~\ref{eqn:posterior}), an
integral over all four of the phase-space distribution parameters
in~$\mvtheta$, can be performed analytically; it leaves
\begin{equation}\label{eq:basic_posterior}
\begin{split}
p(\mvomega | \setofxv) \propto\,\,
  & [\lnepsilon_K - \lnepsilon_L]^{2-N}
    \left[1-[1-e_L]^{2-N}-[e_M]^{2-N} + [e_M-e_L]^{2-N}\right]\\
  & \quad\times\prod_{i=1}^N
    |J(\lnepsilon,e,\angle;r_i,v_{r,i},\vperpi)|\quad ,
\end{split}
\end{equation}
where $\lnepsilon_L$ is the lowest planetary binding energy at this
point $\mvomega$ in dynamical parameter space, $\lnepsilon_K$ is the
highest, $e_L$ is the lowest planetary radial asymmetry at this point
$\mvomega$, $e_M$ is the highest, and we have taken the limit in which
the range of the parameters $(\lnepsilon_a,\lnepsilon_b)$ goes to
infinity, or $A \rightarrow -\infty$, $B \rightarrow \infty$. This
posterior probability distribution for $\mvomega$ represents our
posterior beliefs marginalized over all possible values of
$\{\lnepsilon_a,\lnepsilon_b,e_a,e_b\}$ of the tophat model.

\subsection{Alternative methods}
\label{sec:altmethods}

The results of any inference depend both on the data and on modeling
assumptions. Thus, it is always sensible to question any assumptions
and investigate how sensitive the results are to them.

The marginalization in our basic method is tractable at the expense of
believability: it seems unlikely that the true distribution function
over the planet properties is uniform in the chosen
parameterization. If the true distribution function is far from this
assumption, the inference performed here could be in trouble because
the zero prior probability assigned to all other reasonable forms of
the distribution function makes it impossible for the model to learn
this. This family of distribution functions also has no special
status: a distribution which is uniform in one parameterization will
not be uniform in another. Instead of modeling the distribution in
$(\lnepsilon, e,\angle)$, we could have chosen to model the
distribution function in terms of $(\lnepsilon,e^2,\angle)$. This set
of parameters is, in a sense, more natural as it is closer to
action--angle coordinates, and, thus, the Jacobian
$|J(\lnepsilon,e^2,\angle;r,v_r,\vperp)|$ is more close to
constant. Indeed, from \eqnname~(\ref{eq:jacobian}) it is clear that,
in the case of $\alpha = 2$, this Jacobian does not depend on the
eccentricity of the orbit any longer and
\figurename~\ref{fig:ejacobiansPlanets} shows that this Jacobian is
close to constant for all values of $\ln A$ and $\alpha$. However, it
is not obvious a priori whether using this parameterization will give
better results.  Rather than guessing, we can assume that the
distribution over radial asymmetries is uniform in an unknown
parameterization, and infer from the data what this parameterization
should be. We tried assuming that the radial asymmetry was uniform
over some range of $e$, $e^2$ or $\sqrt{e}$, with equal prior
probability over each choice. This method corresponds to extending
$\mvtheta$ to parameterize a richer class of distributions. As we did
with nuisance parameters before, we can marginalize over this choice.

In general we could marginalize over a very flexible class of
distributions for both the binding energy and radial asymmetry. A
generic way of representing distributions is with a histogram, where
any distribution can be represented within an arbitrary tolerance for
sufficiently narrow bins. Some care is required in putting a prior
$p(\mvtheta)$ over the heights of the bins. It is tempting to use a
Dirichlet distribution, which allows analytical marginalization.
However, this assumes that the heights of the bins are unrelated
except through their overall normalization. An arbitrarily spiky
distribution function is not only unphysical, but we would be unable
to infer its shape from point observations.

We attempted to construct a sensible prior over distributions using
histograms with many bins by coupling the heights of the bins so that
the densities appear smooth. Since the focus on this paper is on the
principles of dynamical inference and not on the specific result on
the Solar System's force law, we only provide a brief description of
how we implemented this approach. We refer the interested reader to
the references provided in the following for more details on the
specifics. A multivariate Gaussian prior can be put over the log bin
heights, approximating a logistic Gaussian process
\citep{leonard1978}. We used 100 equal-width bins coupled with a
Gaussian covariance function \citep{rasmussen2005a}. We put uniform
priors over the start and end points of the histogram and over
physically reasonable ranges of the covariance function's parameters.
The posterior over the dynamical parameters was estimated by Markov
chain Monte Carlo simulation of all the unknowns, Gibbs sampling was
used to update the dynamical parameters over a grid of values for
which we had precomputed the Jacobian term. Slice sampling
\citep{neal2003a} was used to update all nuisance parameters, except
the bin heights which were updated with a Metropolis--Hastings method
\citep[Eq.~15]{neal1999a}.

\section{Results and discussion}

In our basic method we evaluate the Jacobian $|J(\ln
\epsilon,e,\angle;r_i,v_{r,i},\vperpi)|$ for each planet at the
observed radius, radial velocity, and specific angular momentum, for
each value of the dynamical parameters $\mvomega$, and multiply the
magnitudes of the determinants of these Jacobians together. Following
equation (\ref{eq:basic_posterior}) we multiply this product with the
value of the marginalized product of the distribution function and
finally multiply this with the (uniform) prior on the dynamical
parameters in order to obtain the posterior probability distribution
for the dynamical parameters.

\begin{figure}
\includegraphics[height=.2\textheight]{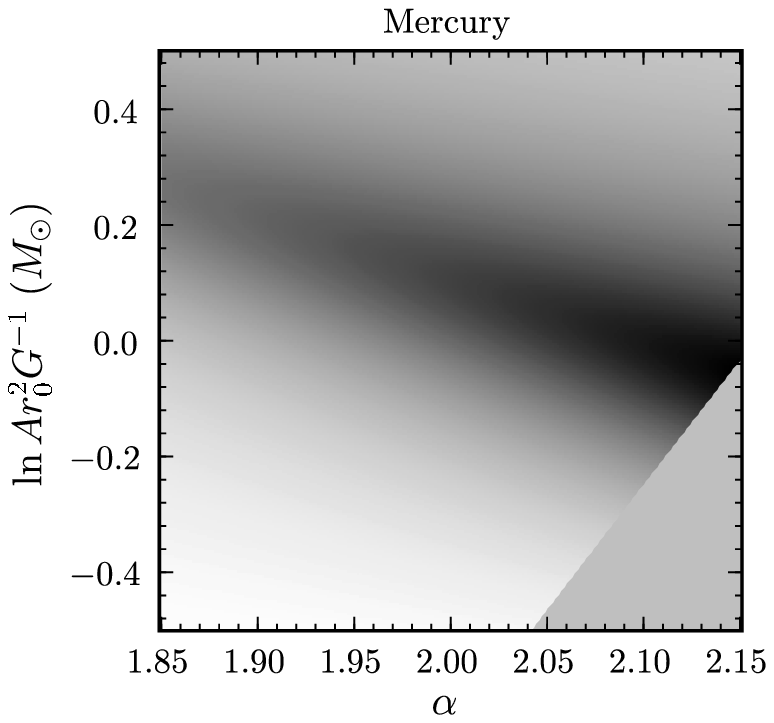}
\includegraphics[height=.2\textheight]{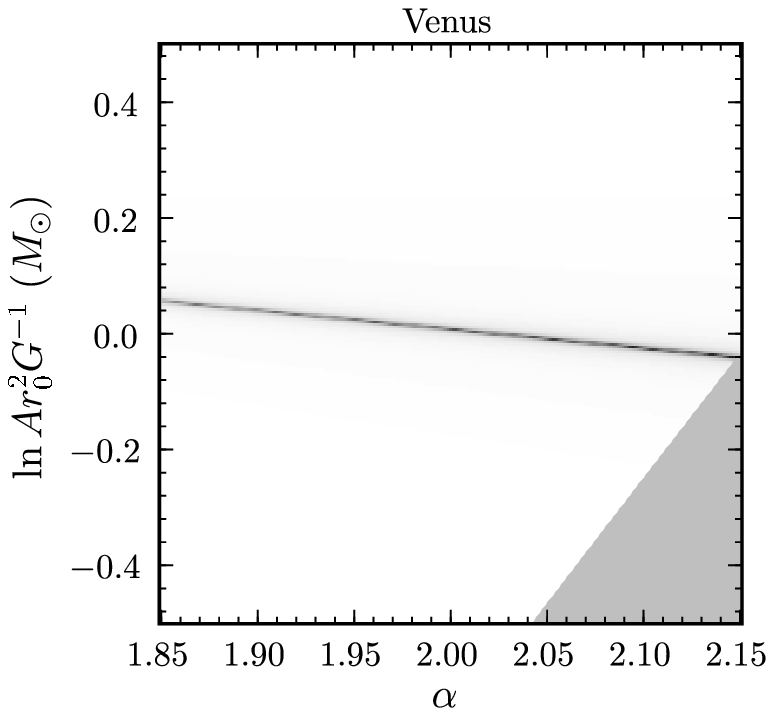}\\[5pt]
\includegraphics[height=.2\textheight]{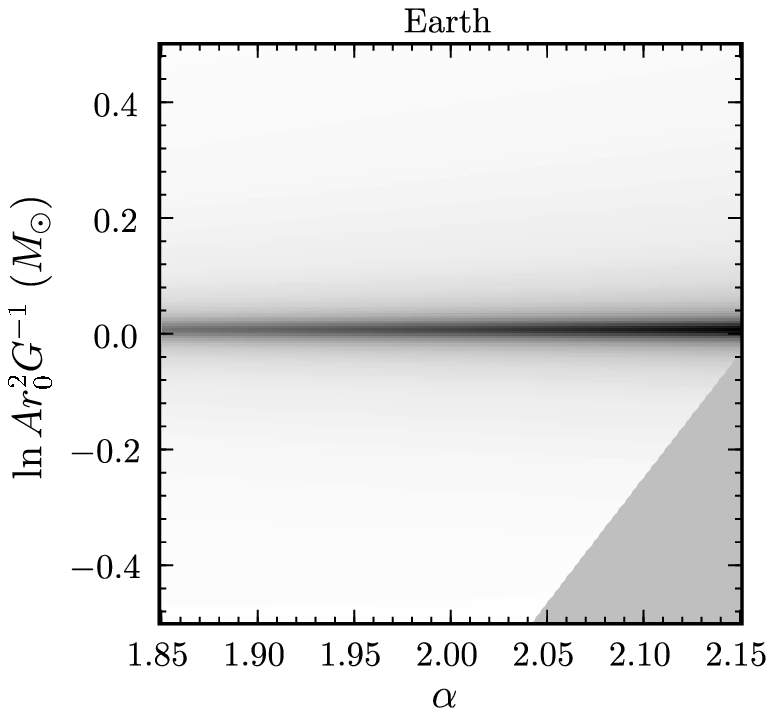}
\includegraphics[height=.2\textheight]{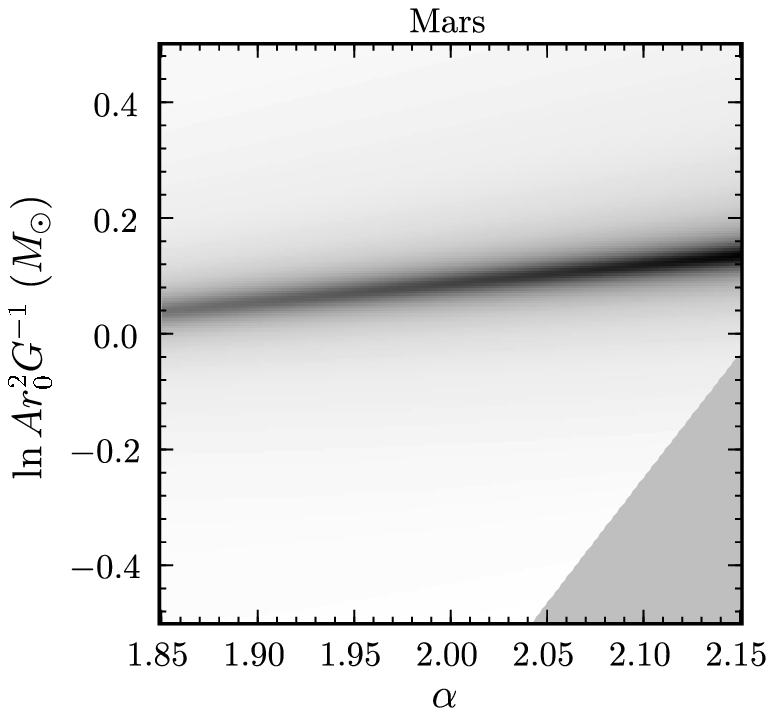}\\[5pt]
\includegraphics[height=.2\textheight]{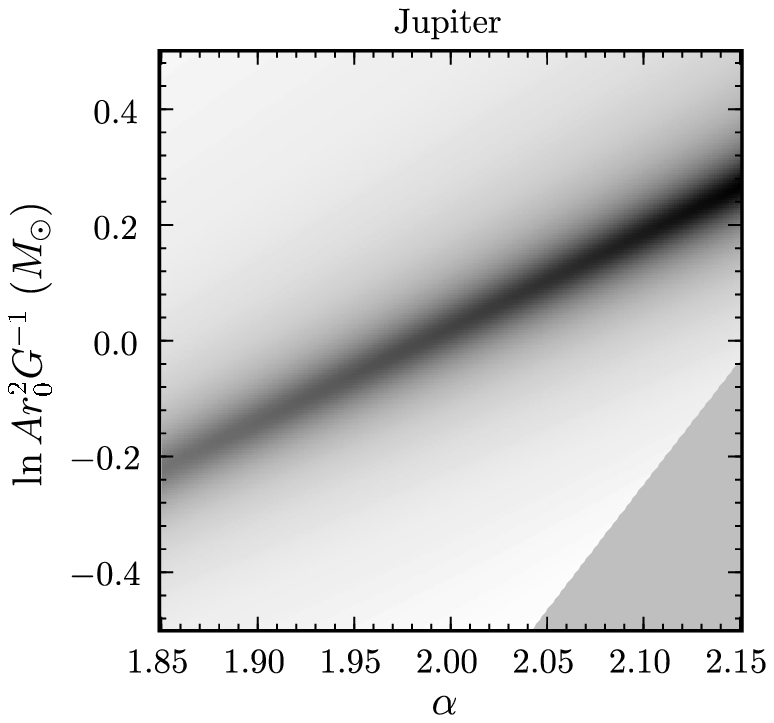}
\includegraphics[height=.2\textheight]{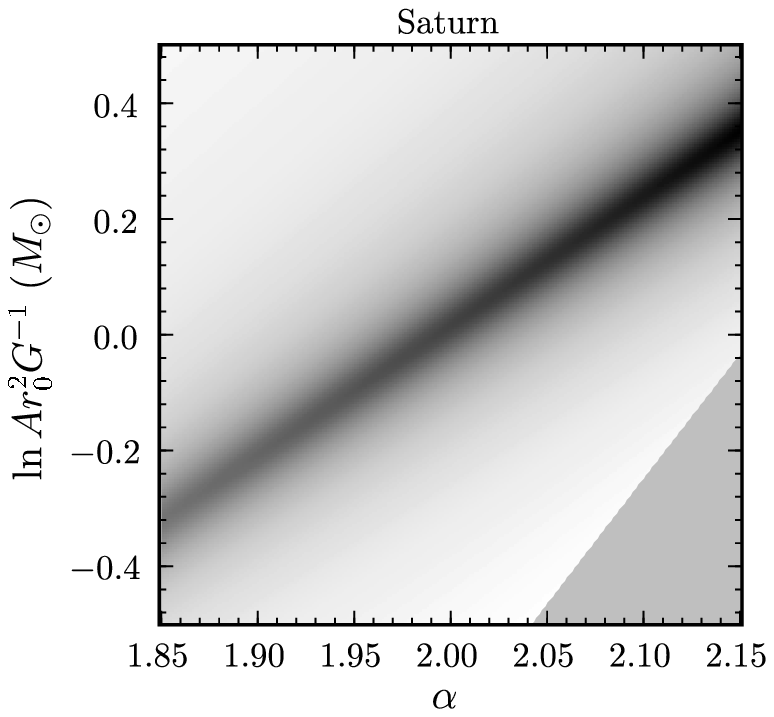}\\[5pt]
\includegraphics[height=.2\textheight]{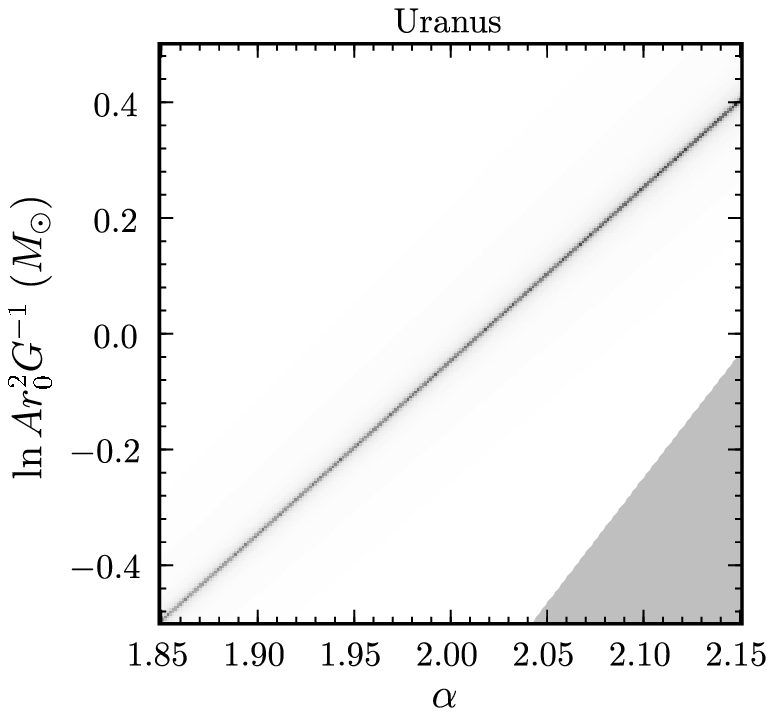}
\includegraphics[height=.2\textheight]{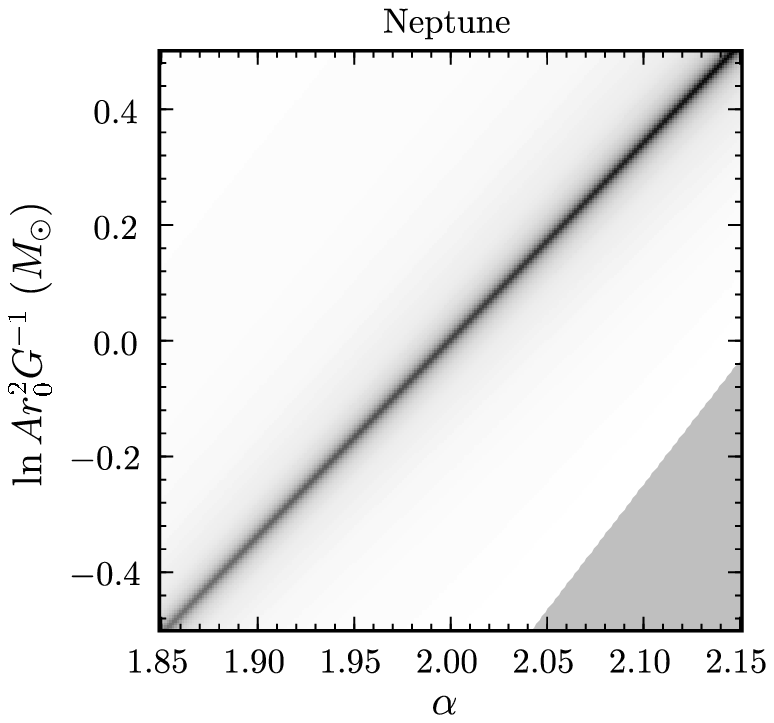}
\caption{Density plots of the absolute value of the determinant
$|J(\lnepsilon,e,\angle;r,v_r,\vperp)|$ of the Jacobian of the
transformation from the energy $\epsilon$, radial asymmetry $e$, and
radial angle $\angle$ coordinates to the relevant positional and
kinematical observables, evaluated at the observed positions and
velocities of the planets, as a function of the dynamical
parameters. Gray scales are linear with darker shades for larger
values.}\label{fig:jacobiansPlanets}
\end{figure}

It is instructive to look at the magnitude of the determinant of the
Jacobian for each planet as a function of the dynamical parameters in
\figurename~\ref{fig:jacobiansPlanets}. These factors seem to show a
clear preference for the dynamical parameters of each planet to lie on
its virial locus. Why is this? For nearly circular orbits the last two
factors in \eqnname~(\ref{eq:dedj2}) are equal and are unbounded for
perfectly circular orbits; this results in a natural preference for
nearly circular orbits for all of the planets, \ie, for each planet to
lie somewhere along its virial locus. At the same time the measured
non-zero radial velocity limits the extent to which a planet's orbit
can be circularized by a suitable choice of the dynamical parameters
and, thus, how close to the circular singularity an individual planet
can be brought. This naturally leads to a weighting scheme in which
planets that are closer to circular orbits (as measured by the
observed angle between the planet's velocity and the tangent to the
circular orbit going through its present location) receive a higher
weight in the analysis than planets on less circular orbits. This
weighting scheme is good in that planets on more circular orbits are
indeed more informative about the potential than planets on less
circular orbits. Therefore, it may seem that the decision to model the
radial asymmetries as uniform in $e$ was fortuitous, but in fact, as
we will discuss below in \sectionname~\ref{sec:sensitivity}, the
results obtained using the basic method are robust in that more
flexible models learn that the tophat in $e$ leads to a good
interpretation of the data.

\begin{figure}
\includegraphics[width=0.6\textwidth]{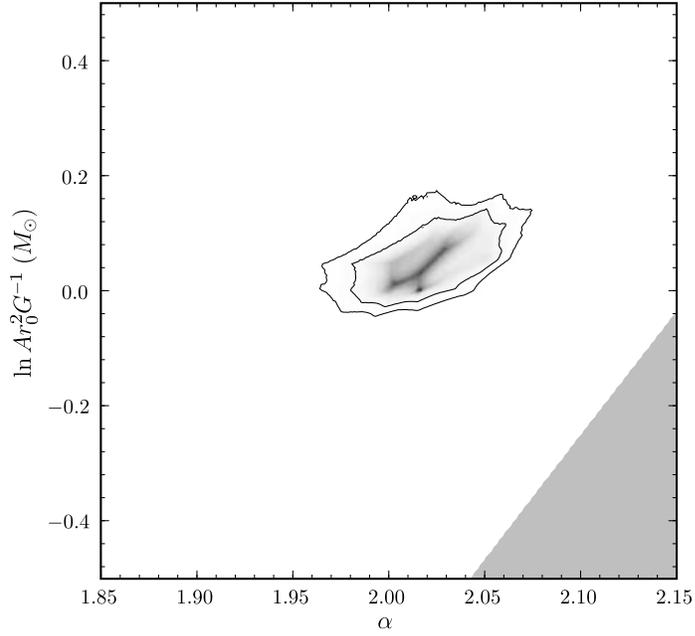}
\caption{The posterior probability distribution $p(\mvomega|\setofxv)$
  for the dynamical parameters on a linear scale. Contours are 95 and
  99~percent posterior regions.}\label{fig:Bayes2d}
\end{figure}

The posterior probability distribution is shown in
\figurename~\ref{fig:Bayes2d}. A strong peak is apparent near $\alpha
= 2$ and the Newtonian Solar value for $\ln A$.  The width of the
probability distribution is indicated by the 95 and 99~percent
posterior contours. These contours are defined to enclose the
smallest area that holds 95 and 99~percent, respectively, of the
posterior probability distribution.

\begin{figure}
\includegraphics[width=0.6\textwidth]{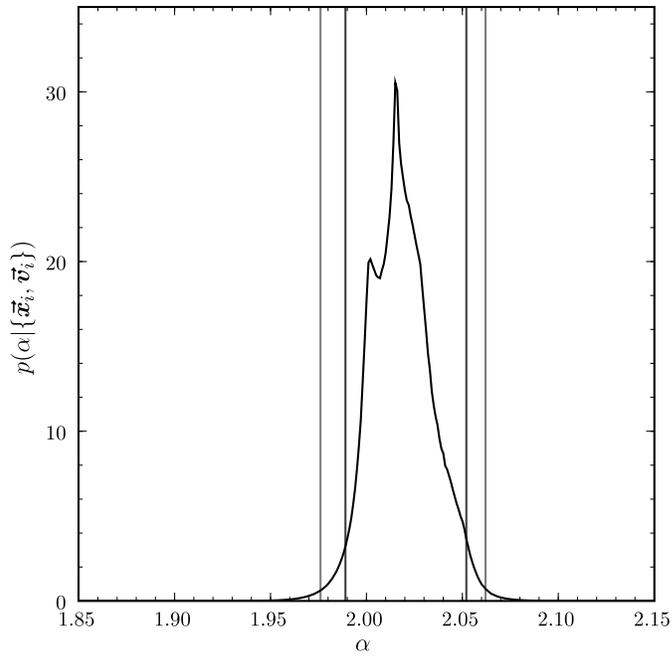}
\caption{Marginalized posterior probability distribution for the
parameter $\alpha$ with 95 and 99~percent posterior
intervals.}\label{fig:Bayes1d}
\end{figure}

In order to infer the exponent $\alpha$ of the force power-law we
perform a second marginalization of the posterior probability
distribution, this time over the (for our purposes) uninteresting
parameter $\ln A$. This gives a posterior probability distribution for
the parameter~$\alpha$
\begin{equation}
p(\alpha|\setofxv) = \int \dd \ln A\, p(\mvomega|\setofxv) \quad.
\end{equation}
This probability distribution is shown in
\figurename~\ref{fig:Bayes1d}. The 95 and 99~percent posterior
intervals in this figure are defined to exclude 2.5 and 0.5~percent,
respectively, of the distribution on either side of the central
region.

The result of the inference is not a value for the parameters but a
posterior probability distribution.  A ``best fit'' value for $\alpha$
could be obtained according to various criteria. For example, the
posterior mean minimizes the expected square difference from the true
parameter. However, since the posterior distribution is multi-modal, a
point-estimate does not capture our uncertainty.  The posterior is
better summarized by a credible interval, containing a given fraction
of the probability mass. The 95~percent posterior interval is $1.989 <
\alpha < 2.052$. This compares favorably with the results obtained by
\citet{Newton} who inferred $\alpha = 2$ from a much richer data set
(though a less rich model set). Modern tests constrain the value of
the exponent $\alpha$ to a fractional accuracy of 10$^{-8}$--10$^{-9}$
on Solar System scales \citep{adelberger,fischbach} using Lunar Laser
Ranging tests \citep{williams} and Keplerian tests comparing $G(r)
M_{\odot}$ and the rate of precession for different planets
\citep{talmadge}.

Our method appears to work well: the true dynamical parameters $A$
and~$\alpha$ are plausible under a fairly tight posterior
distribution, see \figurename~\ref{fig:Bayes2d}. The confidence
intervals from the frequentist tests, \figurename~\ref{fig:freq}, are
somewhat broader than the posterior intervals, but are not directly
comparable. Frequentist confidence intervals do not represent beliefs
about the parameters for this dataset, simply a region that satisfies
some sampling properties. As a result it is unsurprising that the
frequentist confidence intervals appear less good if interpreted as
beliefs given the data. It is difficult to use the frequentist
confidence intervals to make statements about $\alpha$ alone: without
the assumptions of the Bayesian method we cannot marginalize over the
parameter~$A$. One approach that allows frequentist guarantees is to
set $A$ adversarially, but this pessimistic view would give
unreasonably large error bars for~$\alpha$.

That the posterior distribution $p(\alpha|\setofxv)$ turns out to be
multi-modal is no surprise in the light of the virial considerations
from \figurenames~\ref{fig:virial_main} and~\ref{fig:virial_zoom}. The
two peaks in $p(\alpha|\setofxv)$ correspond to the two main
regions in parameter space in which the different virial loci
cross. These virial considerations, and also the frequentist
techniques, are very similar to the probabilistic approach in that
they all prefer each planet to be in a non-special region of radial
angle space, that is, between perihelion and aphelion; this can only
happen simultaneously near the points in parameter space at which
virial loci cross.  The advantage of the probabilistic approach is
that it explains and quantifies this reasoning, and uses it to set
formal limits on the dynamical parameters.

\subsection{Sensitivity analysis}\label{sec:sensitivity}

\begin{figure}
\includegraphics[height=.4\textheight]{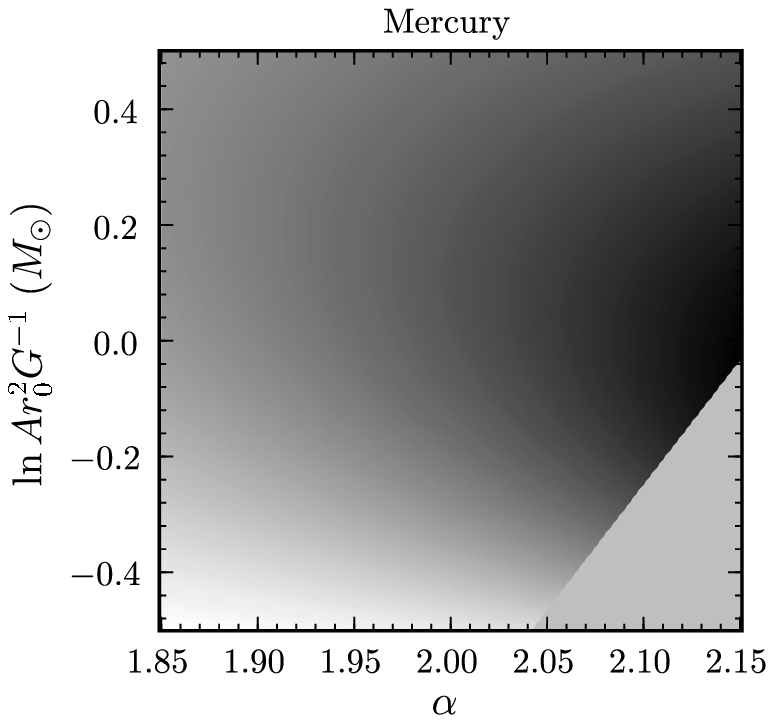}\\[5pt]
\includegraphics[height=.4\textheight]{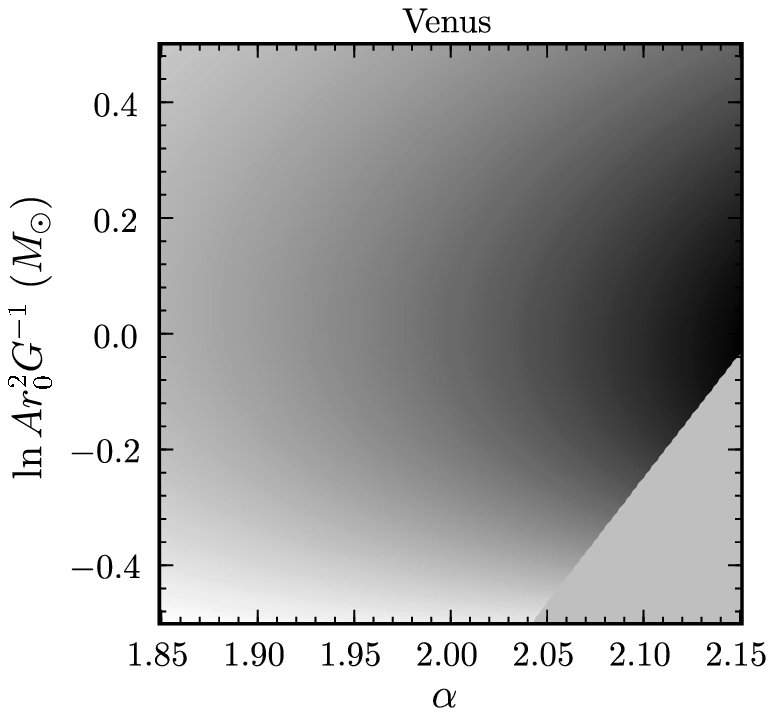}
\caption{Density plots of the absolute value of the determinant
  $|J(\lnepsilon,e^2,\angle;r,v_r,\vperp)|$ of the Jacobian of the
  transformation from the energy $\epsilon$, radial asymmetry squared
  $e^2$, and radial angle $\angle$ coordinates to the relevant
  positional and kinematical observables, evaluated at the observed
  positions and velocities of the planets, as a function of the
  dynamical parameters. Grayscales are linear with darker shades for
  larger values. Only the Jacobians for Mercury and Venus are shown
  here; the corresponding Jacobians for the other planets are very
  similar to these.}\label{fig:ejacobiansPlanets}
\end{figure}

The strongly peaked Jacobian factors shown in
\figurename~\ref{fig:jacobiansPlanets} depend strongly on the choice
of parameterization. If the model was expressed using the radial
asymmetries squared rather than the radial asymmetries, the Jacobians,
some of which are shown in \figurename~\ref{fig:ejacobiansPlanets},
are much flatter. If the model is maintained by also transforming the
prior over radial asymmetry distributions to the new parameterization,
the posterior distribution over dynamical parameters will be
unchanged. Setting all the radial asymmetries close to one is
penalized elsewhere in the mathematical expression.

However, setting the prior over radial asymmetry distributions to
consider only uniform distributions was a strong assumption. If we had
only considered distributions uniform in the radial asymmetry squared,
this combined with the flat Jacobians in
\figurename~\ref{fig:ejacobiansPlanets} would not have given as tight
a posterior distribution. Fortunately we can infer a suitable
parameterization, or equivalently the distribution function from a
richer family of possibilities, from the data.

\begin{figure}
\includegraphics[height=.4\textheight]{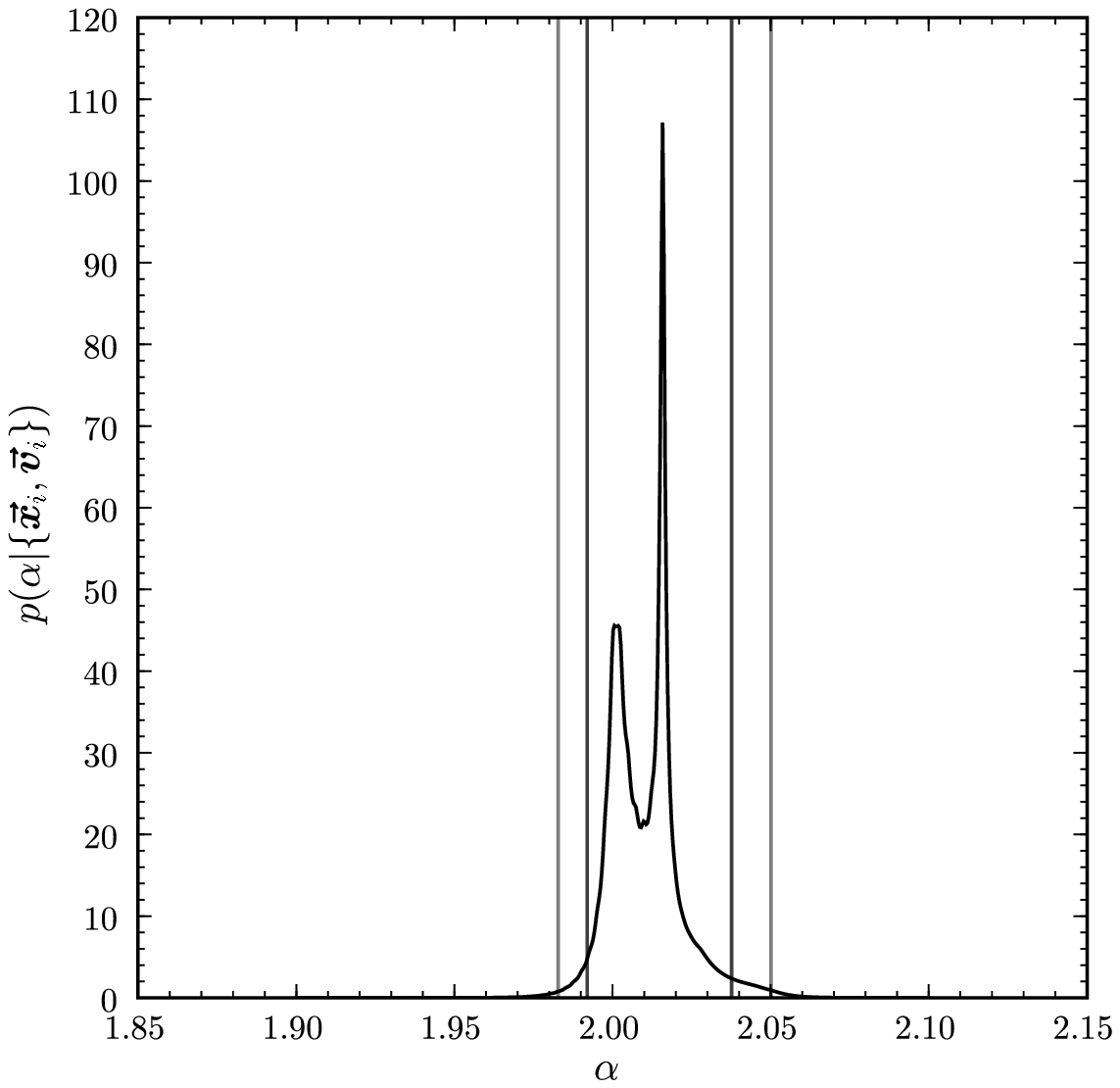}\\
\includegraphics[height=.4\textheight]{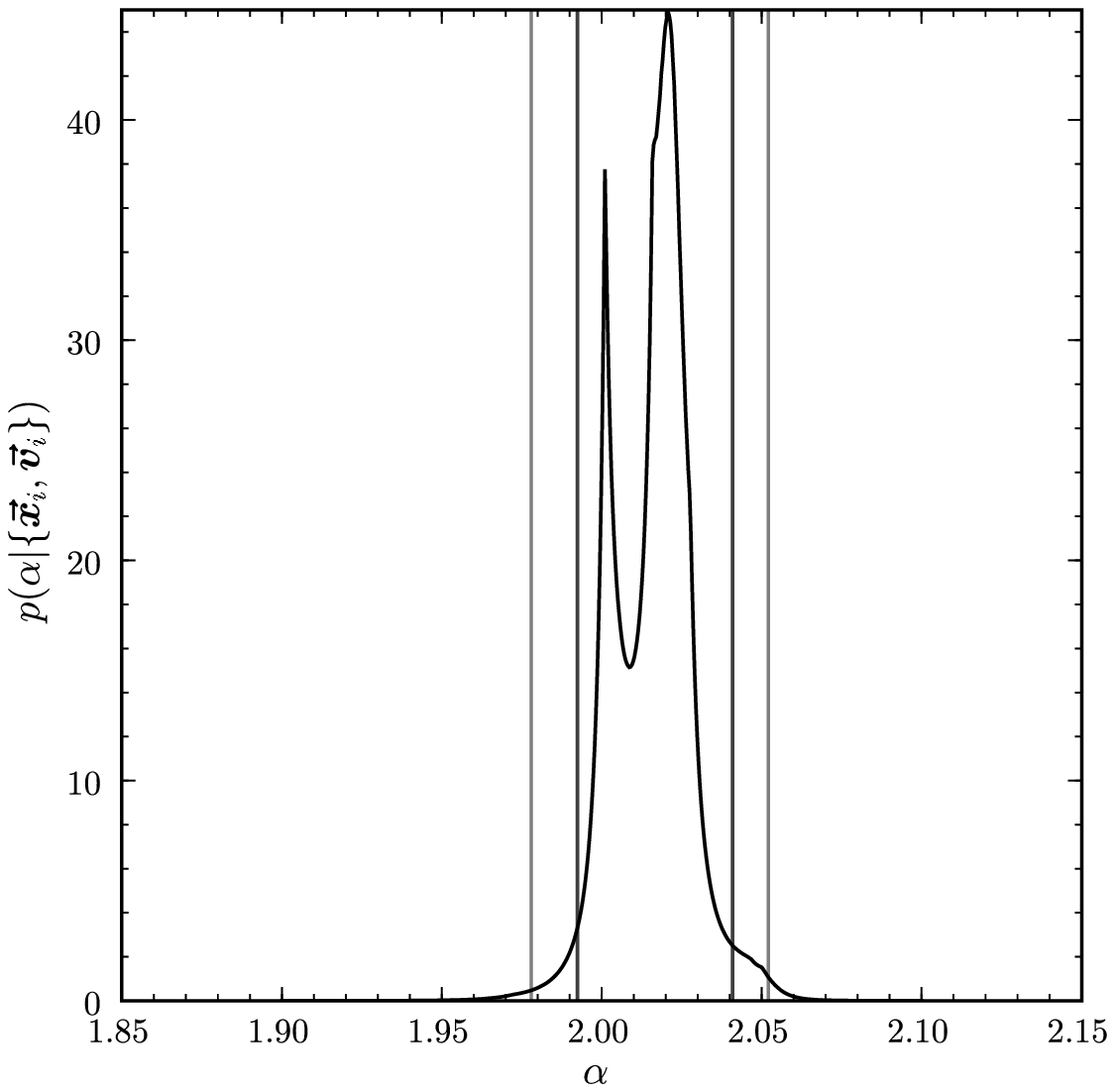}
\caption{Alternative posterior probability distributions for the
parameter $\alpha$ with 95 and 99~percent posterior intervals. Top:
distribution over radial asymmetry is assumed to be uniform in one of
$\sqrt{e}$, $e$ or~$e^2$. Bottom: results from a non-parametric prior
over the distributions of both $e$ and~$\ln A$.}\label{fig:altBayes1d}
\end{figure}

\figurename~\ref{fig:altBayes1d} shows the posterior over $\alpha$
from using the two more flexible priors outlined in
\sectionname~\ref{sec:altmethods}. Assuming that the radial asymmetry is
uniform in one of $\sqrt{e}$, $e$, or~$e^2$ gave a 95\% credible
interval of $1.990 < \alpha < 2.035$. A variety of different flexible
priors for the radial asymmetry distribution gave very similar
results.  Allowing the $\ln A$ constant in the force law to come from
a flexible family of distributions changed the posterior very
little. The 95\% credible interval from the flexible non-parametric
prior on both distributions gave a 95\% credible interval of $1.991 <
\alpha < 2.040$. The fine details of the posterior distribution,
specifically the relative mass in the two modes, are sensitive to
prior assumptions. This is unsurprising with only eight data
points. However, the position of the bulk of the posterior mass, which
gives the confidence interval is surprisingly robust across a wide
range of modeling choices.

As a warning, it is possible to obtain poor results through bad
modeling choices. The reasonable posterior from our basic method,
\figurename~\ref{fig:Bayes1d}, was to some extent due to chance. The
family of possible radial asymmetry distributions is inflexible as it
forces this distribution to be flat over some range in radial
asymmetry. Using the basic method it is not possible to learn the true
distribution, which is not flat in radial asymmetry, from data,
because the truth is assigned zero prior probability. We were
fortunate to choose a parameterization in which the inflexible family
was not too unreasonable. The non-parametric approach is much more
flexible in that its weak smoothness constraints assign non-zero prior
probabilities to all reasonable distribution functions such that the
true one could be picked out with enough data. However, care must also be
taken with non-parametric modeling. An unsmooth Dirichlet prior over
bin heights in a non-parametric prior over distributions gave a broad
and biased distribution over~$\alpha$.

\subsection{Future work}

The necessity of a proper probabilistic approach as laid out in this
paper has been recognized before in the context of inferring the mass
of the Galaxy from a small set of highly informative tracers: distant
satellite galaxies \citep{Little87a} and high velocity stars
\citep{Leonard90a}. The limitations of those works are that they
choose restrictive forms of the distribution function in which the
distribution of angular momentum is fixed as either radial or
isotropic, or given as an inflexible parametric function in between
these two extrema \citep{Kochanek96a}. The precision of their results
for the fundamental parameters of the Galaxy is limited by the
systematic uncertainty arising from this assumption. In the case of
estimating the local escape velocity from a sample of high velocity
stars, a power law model for the energy part of the distribution
function was used; some progress has been made marginalizing over the
exponent, a nuisance parameter, with a prior coming from cosmological
simulations \citep{Smith07a}. In the future, using the technique
introduced in this paper there is no need to make strong assumptions
about the degree of anisotropy of the system at hand or the form of
the distribution function beyond the assumption of complete angle
mixing. The results of this paper show that by introducing flexible
models for the distribution function that can learn from the data,
robust constraints on the parameters of a dynamical system can be
obtained.

The approach developed here can be applied to other (perhaps more
pressing) dynamical inference problems in which test particles can be
relied upon to be well-mixed in angle space. One such problem is the
dynamics in the region surrounding the black hole at the Galactic
Center. Often in these problems complications arise because of large
observational uncertainties (often highly correlated), the absence of
some of the six-dimensional phase-space coordinates, and selection
effects, all of which are absent in the simple problem considered
here.  It will be necessary for these problems to model the full
six-dimensional phase-space distribution function.  This will
complicate the marginalization over the phase-space parameters, which
was trivial here in the basic method, but at the same time it will
permit the discovery of structure in the phase-space distribution,
which can aid with the presence of missing data and large
observational uncertainties.  That this approach has much potential
has been shown before in the case of the Galactic Center; the
assumption that a set of stars at the Galactic Center is part of a
disk-like population has been successful in reconstructing missing
data \citep{Beloborodov06a}. Extension of the approach developed in
this paper will permit incorporation of much more of the available
data on the dynamics in the central region of the Galaxy.  This, in
turn, will lead to a better determination of the mass of the black
hole and the surrounding density profile.

On larger scales, approaches like that developed in this paper will
prove to be essential for the analysis of the large data sets of
upcoming surveys such as \Gaia. As the duration of the \Gaia\ mission
is vanishingly smaller than any dynamical timescale, the problem posed
is essentially the same as the problem posed here: Infer the dynamics
from a snapshot of the kinematics.  Our approach cannot simply be
applied to this larger problem because the system is almost certainly
\emph{not} (trivially) integrable, and the assumption of mixed angles
and lack of resonances is invalidated by the clear abundance of
substructure in the halo \citep[\eg,
][]{willman,belokurovfield,belokurov,koposov} and the disk
\citep{dehnen98b,Bovy09a}; indeed, Jeans himself, in the paper in
which he first wrote down his eponymous theorem \citep{Jeans15a},
argued \emph{against} assuming a steady state for the Galaxy.
Modeling the details of the phase-space distribution function will be
even more important in this context.  However, we expect that a large
fraction of the Galaxy, even a large fraction of the stellar halo, is
well-mixed, such that mixed-angle approaches are expected to lead to
valuable inferences. Combining the mixed-angle and
unmixed-substructure techniques into a general inference about the
dynamics of the Galaxy requires a fully probabilistic method and the
approach developed in this paper is a baby step toward this ambitious
goal.

\acknowledgments First of all we thank Scott Tremaine for many
inspiring discussions that helped shape this paper. It is also a
pleasure to thank Kyle Cranmer, Dustin Lang, Yuri Levin, Phil
Marshall, Bill Press, and Hans-Walter Rix for very helpful discussions
and the anonymous referee for valuable comments. This research was
partially supported by NASA (ADP grant NNX08AJ48G), the NSF (grant
AST-0908357), and a Research Fellowship of the Alexander von Humboldt
Foundation. This project made use of the HORIZONS System provided by
the Solar System Dynamics Group of the Jet Propulsion Laboratory, the
NASA Astrophysics Data System, and the open-source Python modules
scipy, numpy, and matplotlib.

\clearpage
\begin{deluxetable}{lr@{.}lr@{.}lr@{.}lr@{.}lr@{.}lr@{.}l}
\rotate
\tablecaption{Planet Ephemerides for 2009-Apr-01 00:00:00.0000 (CT\tablenotemark{\protect{\ref{CT}}})\label{table:eph}}
\tablecolumns{13}
\tablewidth{0pt}
\tablehead{\colhead{Planet} & \multicolumn{2}{c}{$x$} & \multicolumn{2}{c}{$y$} & \multicolumn{2}{c}{$z$} & \multicolumn{2}{c}{$v_x$} & \multicolumn{2}{c}{$v
_y$} & \multicolumn{2}{c}{$v_z$} \\
\colhead{} & \multicolumn{2}{c}{($\AU$)}& \multicolumn{2}{c}{($\AU$)}& \multicolumn{2}{c}{($\AU$)}& \multicolumn{2}{c}{($\AU$ $\yr^{-1}$)}& \multicolumn{2}{c
}{($\AU$ $\yr^{-1}$)}& \multicolumn{2}{c}{($\AU$ $\yr^{-1}$)}}
\startdata
Mercury & 0 & 324190175 & 0 & 090955208 & -0 & 022920510 & -4 & 627851589 & 10 & 390063716 & 1 & 273504997\\
Venus & -0 & 701534590 & -0 & 168809218 & 0 & 037947785 & 1 & 725066954 & -7 & 205747212 & -0 & 198268558\\
Earth & -0 & 982564148 & -0 & 191145980 & -0 & 000014724 & 1 & 126784520 & -6 & 187988860 & 0 & 000330572\\
Mars & 1 & 104185888 & -0 & 826097003 & -0 & 044595990 & 3 & 260215854 & 4 & 524583075 & 0 & 014760239\\
Jupiter & 3 & 266443877 & -3 & 888055863 & -0 & 057015321 & 2 & 076140727 & 1 & 904040630 & -0 & 054374153\\
Saturn & -9 & 218802228 & 1 & 788299816 & 0 & 335737817 & -0 & 496457364 & -2 & 005021061 & 0 & 054667082\\
Uranus & 19 & 930781147 & -2 & 555241579 & -0 & 267710968 & 0 & 172224285 & 1 & 357933443 & 0 & 002836325\\
Neptune & 24 & 323085642 & -17 & 606227355 & -0 & 197974999 & 0 & 664855006 & 0 & 935497207 & -0 & 034716967\\
\enddata
\setcounter{tableone}{1}
\makeatletter
\let\@currentlabel\oldlabel
\newcommand{\@currentlabel}{\thetableone}
\makeatother
\renewcommand{\thetableone}{\alph{tableone}}
\tablenotetext{\thetableone}{\label{CT}
CT is a coordinate time used in connection with ephemerides. It differs from UTC by about 66 seconds (see \url{http://ssd.jpl.nasa.gov/?horizons_doc\#timesys}).\stepcounter{tableone}}
\tablecomments{The $xyz$-coordinate system is defined as follows: the
  $xy$-plane is given by the plane of the Earth's orbit at J2000.0,
  the $x$-axis is out along the ascending node of the instantaneous
  plane of the Earth's orbit and the Earth's mean equator at J2000.0,
  and the $z$-axis is perpendicular to the $xy$-plane in the
  directional (+ or -) sense of Earth's north pole at J2000.0. The
  origin of the coordinate system is given by the barycenter of the
  Solar System. One year is defined as 365.25 days.}
\end{deluxetable}
\makeatletter
\renewcommand{\@currentlabel}{\oldlabel}
\makeatother

\clearpage

\clearpage

\clearpage

\clearpage

\clearpage

\clearpage

\clearpage

\clearpage

\clearpage


\begin{thebibliography}{}
\bibitem[Adelberger, Heckel, \& Nelson(2003)]{adelberger}
  Adelberger,~E.~G., Heckel,~B.~R., \& Nelson,~A.~E., 2003,
  Annu.~Rev.~Nucl.~Part.~Sci., 53, 77
\bibitem[Anderson \& Darling(1952)]{AndersonDarling}
  Anderson,~T.~W. \& Darling,~D.~A., 1952,
  Ann.\,Math.\,Stat., 23, 193
\bibitem[Beloborodov \& Levin(2004)]{roulette}
  Beloborodov,~A.~M. \& Levin,~Y., 2004,
  \apj, 613, 224
\bibitem[Beloborodov \etal(2006)]{Beloborodov06a} 
  Beloborodov,~A.~M, Levin,~Y., Eisenhauer,~F., Genzel,~R., 
  Paumard,~T., Gillessen,~S.,
  \& Ott,~T., 2004,
  \apj, 648, 405
\bibitem[Belokurov \etal(2006)]{belokurovfield}
  Belokurov,~V., \etal, 2006,
  \apjl, 642, L137
\bibitem[Belokurov \etal(2007)]{belokurov}
  Belokurov,~V., \etal, 2007,
  \apj, 654, 897
\bibitem[Binney \& Tremaine(2008)]{binneytremaine}
  Binney,~J. \& Tremaine,~S., 2008, Galactic Dynamics: Second Edition
  (Princeton University Press)
\bibitem[Bovy, Hogg, \& Roweis(2009)]{Bovy09a}
  Bovy,~J., Hogg,~D.~W., \& Roweis,~S.~T., 2009,
  \apj, 700, 1794
\bibitem[Dehnen(1998)]{dehnen98b}
  Dehnen,~W., 1998,
  \aj, 115, 2384
\bibitem[Fischbach \& Talmadge(1999)]{fischbach}
  Fischbach,~E. \& Talmadge,~C.~L., 1999
  The Search for Non-Newtonian Gravity,
  (Berlin: Springer)
\bibitem[Giorgini \etal(1996)]{Giorgini96a}
  Giorgini,~J.~D., Yeomans,~D.~K., Chamberlin,~A.~B., Chodas,~P.~W., Jacobson,~R.~A., Keesey,~M.~S., Lieske,~J.~H., Ostro,~S.~J., Standish,~E.~M., \& Wimberly,~R.~N., 1996, 
  \baas, 28, 1158
\bibitem[Jeans(1915)]{Jeans15a}
  Jeans,~J.~H., 1915,
  \mnras, 76, 70
\bibitem[Jeffreys(1939)]{Jeffreys39a}
  Jeffreys,~H., 1939,
  Theory of Probability (Oxford: Clarendon Press)
\bibitem[Johnston \etal(1999)]{Johnston99}
  Johnston,~K.~V., Zhao,~H., Spergel,~D.~N., \& Hernquist, L., 1999,
  \apjl, 512, L109 
\bibitem[Kaasalainen \& Binney(1994)]{Binney94}
  Kaasalainen,~M. \& Binney,~J., 1994,
  Phys.\,Rev.\,Lett., 73, 2377
\bibitem[Kendall(1938)]{Kendall38a}
  Kendall,~M., 1938,
  Biometrika, 30, 81
\bibitem[Kepler(1609)]{Kepler}
  Kepler,~J., 1609,
  Astronomia Nova,
  trans.\ W.~H.~Donahue, 1992
  (Cambridge University Press)
\bibitem[Kochanek(1996)]{Kochanek96a}
  Kochanek,~C.~S., 1996,
  \apj, 457, 228
\bibitem[Kolmogorov(1941)]{Kolmogorov41a}
  Kolmogorov,~A.~N., 1941,
  Ann.~Math.~Stat., 12, 461
\bibitem[Koposov \etal(2008)]{koposov}
  Koposov,~S., \etal, 2008,
  \apj, 686, 279
\bibitem[Kuiper(1962)]{Kuiper62a}
  Kuiper,~N.~H., 1962,
  Proc.~Koninklijke Nederlandse Akademie van Wetenschappen A, 63, 38
\bibitem[Leonard(1978)]{leonard1978}
  Leonard.,~T., 1978,
  J.\,Roy.\,Stat.\,Soc.\,B, 40, 113
\bibitem[Leonard \& Tremaine(1990)]{Leonard90a}
  Leonard,~P.~J.~T. \& Tremaine,~S., 1990,
  \apj, 353, 486
\bibitem[Little \& Tremaine(1987)]{Little87a}
  Little,~B.~\& Tremaine,~S., 1987,
  \apj, 320, 493
\bibitem[Neal(1993)]{neal1993}
  Neal.,~R.~M., 1993,
  Probabilistic Inference Using {M}arkov Chain {M}onte {C}arlo Methods.
  Technical Report CRG-TR-93-1, Department of Computer Science,
  University of Toronto
\bibitem[Neal(1999)]{neal1999a}
  Neal.,~R.~M., 1999,
  in Bayesian Statistics 6,
  ed.\,J.~M. Bernardo \etal. (Oxford University Press), 475
\bibitem[Neal(2003)]{neal2003a}
  Neal.,~R.~M., 2003,
  Ann.\,Stat., 31, 705
\bibitem[Newton(1687)]{Newton}
  Newton,~I., 1687,
  Philosophiae Naturalis Principia Mathematica,
  trans.\ F.~Cajori, 1934
  (University of California Press)
\bibitem[Oort(1932)]{Oort32}
  Oort,~J.~H., 1932,
  Bull.\,Astron.\,Inst.\,Netherlands, 6, 249
\bibitem[Perryman \etal(2001)]{Perryman01a}
  Perryman,~M.~A.~C., \etal, 2001,
  \aap, 369, 339
\bibitem[Press \etal(2007)]{Press07a}
  Press,~W.~H., Teukolsky,~S.~A, Vetterling,~W.~T., \& Flannery,~B.~P., 2007,
  Numerical Recipes: The Art of Scientific Computing, 3rd Edition (Cambridge University Press)
\bibitem[Rasmussen and Williams(2006)]{rasmussen2005a}
  Rasmussen,~C.~E. \& Williams.,~C.~K.~I., 2006
  Gaussian Processes for Machine Learning (MIT Press)
\bibitem[Schwarzschild(1979)]{Schwarzschild79}
  Schwarzschild,~M., 1979
  \apj, 232, 236
\bibitem[Sivia \& Skilling(2006)]{Sivia06a}
  Sivia,~D.~S.~\& Skilling,~J., 2006,
  Data Analysis:\ A Bayesian Tutorial (Oxford University Press)
\bibitem[Smith \etal(2007)]{Smith07a}
  Smith,~M.~C., \etal, 2007,
  \mnras, 379, 755
\bibitem[Standish(1998)]{Standish98a}
  Standish,~E.~M., 1998,
  JPL Planetary and Lunar Ephemerides, DE405/LE405,
  JPL IOM 312, F-98-048 (Pasadena, CA: JPL)
\bibitem[Standish(2004)]{Standish04a}
  Standish,~E.~M., 2004,
  \aap, 417, 1165
\bibitem[Stephens(1970)]{Stephens70a}
  Stephens,~M.~A., 1970,
  J.~R.~Stat.~Soc.~B, 32, 115
\bibitem[Talmadge \etal(1988)]{talmadge}
  Talmadge,~C., Berthias,~J.-P., Hellings,~R.~W., \& Standish,~E.~M., 1988,
  Phys.~Rev.~Lett., 61, 1159
\bibitem[Williams, Turyshev, \& Boggs(2004)]{williams}
  Williams,~J.~G., Turyshev,~S.~G., \& Boggs,~D.~H., 2004,
  Phys.~Rev.~Lett., 93, 261101
\bibitem[Willman \etal(2005)]{willman}
  Willman,~B., \etal, 2005,
  \apjl, 626, L85
\end{thebibliography}
\end{document}